\documentclass[prx,reprint,superscriptaddress]{revtex4-2}
\usepackage[utf8]{inputenc}
\usepackage[export]{adjustbox}
\usepackage{lipsum}
\usepackage{fullpage} % Package to use full page
\usepackage{tikz} % Package for drawing
\usepackage{amsmath,amsfonts,amsthm, mathtools}
\usepackage{dsfont}
\usepackage{csquotes}
\usepackage{amssymb}
\usepackage{verbatim}
\usepackage{bm}

\usepackage[caption=false]{subfig}

\usepackage{hyperref}
\definecolor{darkred}  {rgb}{0.5,0,0}
\definecolor{darkblue} {rgb}{0,0,0.5}
\definecolor{darkgreen}{rgb}{0,0.5,0}
\hypersetup{
  colorlinks = true,
  urlcolor  = blue,         % color of external links
  linkcolor = darkblue,     % color of internal links
  citecolor = darkgreen,    % color of links to bibliography
  filecolor = darkred       % color of file links
}

\theoremstyle{definition}

\newtheorem{definition}{Definition}

\newtheorem{theorem}{Theorem}
\newtheorem*{theorem*}{Theorem}
\newtheorem{lemma}[theorem]{Lemma}
\newtheorem*{lemma*}{Lemma}
\newtheorem{proposition}[theorem]{Proposition}
\newtheorem*{proposition*}{Proposition}

\newtheorem*{remark}{Remark}

\makeatletter
\newtheorem*{rep@theorem}{\rep@title}
\newcommand{\newreptheorem}[2]{%
\newenvironment{rep#1}[1]{%
 \def\rep@title{#2 \ref{##1}}%
 \begin{rep@theorem}}%
 {\end{rep@theorem}}}
\makeatother
\newreptheorem{theorem}{Theorem}
\newreptheorem{lemma}{Lemma}
\newreptheorem{proposition}{Proposition}

\newcommand{\mbf}{\mathbf}
\newcommand{\mbb}{\mathbb}
\newcommand{\mc}{\mathcal}
\newcommand{\msf}{\mathsf}
\newcommand{\N}{\mathcal{N}}
\newcommand{\tr}{\textrm{Tr}}
\newcommand{\wt}{\widetilde}

\newcommand{\cl}{\text{cl}}

\renewcommand{\ol}{\overline}
\usepackage{multirow}
\newcommand{\ket}[1]{|#1\rangle}
\newcommand{\bra}[1]{\langle #1|}
\newcommand{\op}[2]{|#1\rangle\langle #2|}

\newcommand{\state}[1]{\ket{#1}\bra{#1}}

\newcommand{\ass}{\text{ass}}

\definecolor{cool_green}{rgb}{0.0, 0.5, 0.0}

\newcommand{\todo}[1]{{\color{red} #1}}

\vspace{1cm}

\begin{document}

\title{Information Carried by a Single Particle in Quantum Multiple-Access Channels}

\author{Xinan Chen}
\thanks{X. Chen and Y. Zhang contributed equally to this paper.}
\affiliation{Department of Electrical and Computer Engineering, Coordinated Science Laboratory, University of Illinois at Urbana-Champaign, Urbana, IL 61801, USA}
\author{Yujie Zhang}
\thanks{X. Chen and Y. Zhang contributed equally to this paper.}
\affiliation{Department of Physics, University of Illinois at Urbana-Champaign, Urbana, IL 61801, USA}
\author{Andreas Winter}
\affiliation{Institució Catalana de Recerca i Estudis Avançats (ICREA),
Pg. Llu\'{i}s Companys, 23, 08010 Barcelona, Spain}
\affiliation{Grup d’Informació Quàntica, Departament de Física,
Universitat Autònoma de Barcelona, 08193 Bellaterra (Barcelona), Spain}
\affiliation{Institute for Advanced Study, Technische Universit\"at M\"unchen, 
 Lichtenbergstra{\ss}e 2a, D-85748 Garching, Germany}
\author{Virginia O. Lorenz}
\affiliation{Department of Physics, University of Illinois at Urbana-Champaign, Urbana, IL 61801, USA}
\author{Eric Chitambar}
\affiliation{Department of Electrical and Computer Engineering, Coordinated Science Laboratory, University of Illinois at Urbana-Champaign, Urbana, IL 61801, USA}
\email[Email at]{echitamb@illinois.edu}

\date{\today}

\begin{abstract}
  Non-classical features of quantum systems have the potential to strengthen the way we currently exchange information.  In this paper, we explore this enhancement on the most basic level of single particles.  To be more precise, we compare how well multi-party information can be transmitted to a single receiver using just one classical or quantum particle. Our approach is based on a multiple-access communication model in which messages can be encoded into a single particle that is coherently distributed across multiple spatial modes. Theoretically, we derive lower bounds on the accessible information in the quantum setting that strictly separate it from the classical scenario.  This separation is found whenever there is more than one sender, and also when there is just a single sender who has a shared phase reference with the receiver.  Experimentally, we demonstrate such quantum advantage in single-particle communication by implementing a multi-port interferometer with messages being encoded along the different trajectories.  Specifically, we consider a two-sender communication protocol built by a three-port optical interferometer.  In this scenario, the rate sum achievable with a classical particle is upper bounded by one bit, while we experimentally observe a rate sum of $1.0152\pm0.0034$ bits in the quantum setup.
  
  %${R}_{\text{cl}}=1$ while a rate sum of ${R}_{\text{id}}=1.0875$ could be achieved in the ideal quantum case and a rate sum of ${R}_{\text{exp}}=1.0152\pm0.0034$ is observed experimentally. 
\end{abstract}

\maketitle

\section{Introduction}
It is well known that a quantum particle exhibits fundamentally different properties than its classical counterpart. For instance, while a classical particle has a definite trajectory in space, a quantum particle can be placed in a coherent superposition of different paths as it moves from one point in space to another.  A natural practical question is whether this superposition of trajectories can be utilized for performing some communication task \cite{Guerin-2016, Chiribella-2018, Chiribella-2019a, Horvat-2019b, Kristjansson-2020b, Horvat-2021a}.  In this paper, we focus on whether the path coherence of a single particle can be used to enhance the communication of $N$ spatially separated parties to a single receiver. %In other words, can a quantum particle carry more information from a set of distributed senders to one receiver?

Several previous papers have addressed similar questions in this direction. Inspired by the famous two-slit experiment, Massar first showed the advantage of quantum particles in the bipartite fingerprinting task \cite{Massar-2005a}. In such a task, Alice and Bob each possesses one bit $x,y\in\{0,1\}$, and they wish to let a referee decide whether $x=y$ by sending minimal amount of information to the referee.  It is not difficult to see that one quantum particle in the state $1/\sqrt{2}(\ket{0}_\msf{A}\ket{1}_\msf{B}+\ket{1}_\msf{A}\ket{0}_\msf{B})$ suffices for this objective, while in the classical regime, the parties must send both $x$ and $y$ for the referee to certify that $x=y$. In Ref. \cite{DelSanto-2018a}, the authors reinterpreted this result as two-way communication using only one single quantum particle, which is forbidden if the information medium is a classical particle. This idea was further extended to the scenario where Alice and Bob each have an $n$-bit string \cite{Hsu-2020}. Using an $n$-level Mach-Zehnder interferometer, one of Alice and Bob can retrieve the other's full $n$-bit string, while only one bit of information is revealed to the other party. Since this can be done for an arbitrary $n$, this result suggests, roughly speaking, that a single quantum particle can carry an arbitrarily large amount of information in point-to-point communication.  Complementing the point-to-point communication results, it was recently discovered via convex polytope analysis that using a single quantum particle, one can generate multiple-access channels (MACs) that cannot be constructed with a classical particle \cite{Horvat-2021a, Zhang-2022}. However, these latter results pertain to the specific transition probabilities $p(y|x_1,\cdots, x_N)$ of the generated $N$-party MACs.  It has remained elusive whether the discovered non-classical MACs actually have advantages in terms of more practical figures of merit, such as asymptotic communication rates.

In this paper, we provide a positive answer to this question. Specifically, we utilize the framework of single-particle multiple-access channels (MACs) developed in \cite{Zhang-2022} to investigate the achievable rate regions of distributed communication using a single particle. While the communication rate sum of the different senders is always upper bounded by 1 bit if a single classical particle is used, in the quantum setting a rate sum of at least 1.10 bits is achievable for two senders. Even higher rates can be achieved if there are more than two senders. Moreover, we experimentally demonstrate the quantum advantages by implementing one of our designed protocols.  In particular, we achieve a quantum advantage within five standard deviations  using linear optics and a single photon state. \par

This paper is organized as follows. In Section \ref{sect:framework}, we introduce the operational framework of single-particle MACs and review some information-theoretic concepts such as the achievable rate regions of MACs. In Section \ref{sect:theoretical-results}, we study in detail the theoretical aspects of our work. In Section \ref{Sect:experiment}, we give our experimental demonstration of the two-sender coherent assisted communication protocol using linear optics and a heralded single photon state, where quantum-enhanced communication is achieved by preparing a single photon in a superposition of different trajectories. %\todo{(add a table of notation )}

\section{Operational framework and information theory preliminaries}
\label{sect:framework}
\subsection{MACs constructed with one particle}
\label{sect:one-particle-macs}
To compare how much classical information can be carried by a classical or quantum particle with none of its internal
degrees of freedom being accessible, we utilize the framework of single-particle MACs developed in Ref. \cite{Zhang-2022}. This framework, which we now briefly describe, was inspired by previous work \cite{Coles-2016a,Biswas-2017a} that captured the resource-theoretic features of quantum coherence in a multi-port interferometer setup. We denote the collection of senders as $\bm{\msf{A}} = (\msf{A}_1,\msf{A}_2,\cdots,\msf{A}_N)$ and assume that each message sent by each sender is finite. These $N$ senders will use a one-particle state to send information to a single receiver $\msf{B}$. Recall that the Fock space is described by $\mc{H} = \bigoplus_{i=0}^\infty \mc{H}_i$,
where $\mc{H}_i$ is the $i$-particle subspace of $\mc{H}$. A one-particle state is represented by a density operator $\rho^{\bm{\msf{A}}}$ acting on the one-particle subspace, which is
\begin{align}
    \mc{H}_1^{\bm{\msf{A}}} \coloneqq \text{span}\left\{\ket{\mbf{e}_i}:1 \leq i \leq N\right\},
\end{align}
where $\ket{\mbf{e}_i} = \ket{0}^\msf{A_1}\cdots\ket{1}^{\msf{A}_i}\cdots\ket{0}^{\msf{A}_N}$ is the state of the particle on path $i$, with $\ket{0}$ being the vacuum state. The senders then encode their messages using completely positive trace-preserving (CPTP) maps.  For example if party $\msf{A}_i$ wishes to send message $x_i$, the CPTP map $\mc{E}_{x_i}^{\msf{A}_i}$ is locally applied.  The fully encoded state for joint message $\bm{x}\coloneqq(x_1,\cdots,x_N)$ is given by
\begin{align}
    \sigma_{\bm{x}} \coloneqq \sigma_{x_1\cdots x_N} = \mc{E}_{x_1}^{\msf{A}_1}\otimes\cdots\otimes\mc{E}_{x_N}^{\msf{A}_N} (\rho^{\bm{\msf{A}}}).
\end{align}
For the purposes of this investigation, we restrict the allowed CPTP maps that the senders use to encode. Specifically, since we are interested in the information-carrying ability of a single particle, we have to require that the encoding operations cannot increase particle number. More specifically, we model the encoding operations as CPTP maps with a particle number-preserving unitary extension, that is, 
\begin{align}
    \mc{E}(\rho^{\msf{A}}) = \tr_\msf{E}\left[U(\rho^\msf{A}\otimes\state{0}^{\msf{E}})U^\dagger\right],
\end{align}
where $U$ preserves the overall particle number in the system $\msf{A}$ and the environment $\msf{E}$. 
\begin{comment}
, i.e.,
\begin{gather}
    U = U_0 \oplus U_1; \\
    U_0\ket{0}^\msf{AE} = \ket{0}^\msf{AE},\qquad U_1\ket{\psi}^\msf{AE}\in\mc{H}_1 \;\;\forall\ket{\psi}^\msf{AE}\in\mc{H}_1
\end{gather}
\end{comment}
This set of operations was termed \textit{number-preserving extendible (NPE) operations} in Ref. \cite{Zhang-2022} and was fully characterized for an arbitrary number of particles. Here we focus only on the case where there is at most one particle. In this case, these operations are convex combinations of channels with Kraus operators
\begin{align}
    K_1 = 
    \begin{pmatrix}
        1 & 0 \\
        0 & e^{i\phi_1}\sqrt{1-\gamma}
    \end{pmatrix},\qquad
    K_2 = 
    \begin{pmatrix}
        0 & e^{i\phi_2}\sqrt{\gamma} \\
        0 & 0
    \end{pmatrix}.
    \label{eq:NPE-Kraus-operators}
\end{align}
Note they can be seen as generalized amplitude damping channels with two additional relative phase parameters. In this work, we will rely heavily on two particular NPE operations in the encoding: the completely damping operation $\rho\mapsto\mc{E}^{(\text{vac})}(\rho):=\tr(\rho)\state{0}$ and the phase shift operation $\rho\mapsto \mc{E}^{(\phi)}(\rho):= e^{-iZ \phi/ 2}\rho e^{iZ\phi/2}$, where $Z=\left(\begin{smallmatrix}1&0\\0&-1\end{smallmatrix}\right)$.  Note that $\mc{E}^{(\text{vac})}$ and $\mc{E}^{(\phi)}$ correspond to the choices of $\gamma=1$ and $\gamma=0$ in Eq. \eqref{eq:NPE-Kraus-operators}, respectively.  In optical communication, these encoding operations correspond to on-off keying (OOK) modulation and phase-shift keying (PSK) modulation \cite{kato-1999, guha-2008}.

\begin{figure}[t]
\centering
  \includegraphics[width=.45\textwidth]{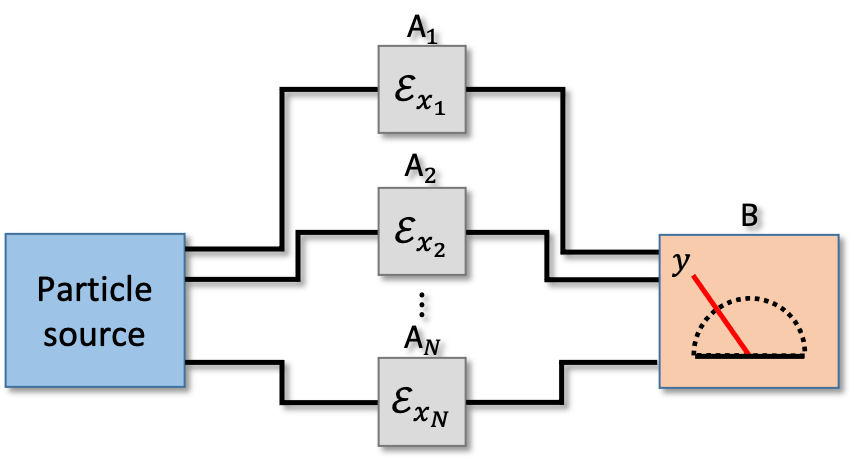}
  \caption{The general scheme for building a multiple-access classical channel using a single particle.}
  \label{fig:non-coherence-assisted-diagram}
\end{figure}

After the encoding operations, the state $\sigma_{\bm{x}}$ is sent to the receiver, and the receiver tries to reconstruct the message using a positive operator-valued measure (POVM) $\{\Pi_y\}_y$. This process induces a classical channel by
\begin{align}
    p(y|\bm{x}) \coloneqq \tr(\Pi_y\sigma_{\bm{x}}).
    \label{eq:mac-def}
\end{align}
A graphical representation of this framework is shown in Fig. \ref{fig:non-coherence-assisted-diagram}. Note that in our model we will always assume that the receiver shares a phase reference with the particle source, and so $\ket{\mbf{e}_i}$ is defined with the same overall phase for both the source and detector \cite{Bartlett-2007}.

%The achievable rates of this channel will be considered as the achievable rates of single-particle communication. A graphical representation of this framework is shown in Fig. \ref{fig:non-coherence-assisted-diagram}. %Here have assumed that, if we send in total $n$ particles individually using our framework, the receiver measures each particle individually. This is a sensible assumption since performing joint measurement on multiple photons is notoriously difficult \cite{}.  However, we can incorporate joint measurements into our framework just as well. This scenario is in fact precisely what we will consider in Sect. \ref{sect:upper-bounds}.  \todo{[[Let's discuss these last few sentences...]]}

With this operational framework in mind, we can define the set of $N$-sender MACs constructed from a classical particle as MACs $p(y|x_1\cdots x_N)$ of the form
\begin{align}
    p(y|x_1\cdots x_N) = \!\tr\left(\Pi_y\left[\mc{E}_{x_1}^{\msf{A}_1}\otimes\cdots\otimes\mc{E}_{x_N}^{\msf{A}_N} (\rho^{\bm{\msf{A}}}_{\cl})\right]\right)
    \label{Eq:classical-Born}
\end{align}
where $\rho^{\bm{\msf{A}}}_\cl = \sum_{i=1}^N p_i\state{\mbf{e}_i}$. In other words, these are MACs that can be constructed from a classical source state that has no coherence between any two paths. On the other hand, in the truly quantum case, no restriction is placed on the initial one-particle density matrix.  We say that the set of $N$-sender MACs constructed from a quantum particle are MACs that have the form
\begin{align}
    p(y|x_1\cdots x_N) =\! \tr\left(\Pi_y\left[\mc{E}_{x_1}^{\msf{A}_1}\otimes\cdots\otimes\mc{E}_{x_N}^{\msf{A}_N} (\rho^{\bm{\msf{A}}})\right]\right)
    \label{Eq:quantum-Born}
\end{align}
where $\rho^{\bm{\msf{A}}}\in\mc{D}(\mc{H}_1)$.  Here $\mc{D}(\mc{H}_1)$ denotes the set of density operators on the one-particle subspace. Throughout this work we assume that the message $x_i$ of party $\msf{A}_i$ is chosen from alphabet set $\mc{X}_i$, which will always be a finite set of integers $\mc{X}_i=[m_i]:=\{0,\cdots,m_i-1\}$.  Similarly, we let $\mc{Y}$ denote the output alphabet of the receiver $\msf{B}$. For input and output alphabet $\bm{\mc{X}}\coloneqq\mc{X}_1\times\mc{X}_2\times\cdots\times\mc{X}_N$ and $\mc{Y}$, we denote the set of classical MACs by $\mc{C}_N(\bm{\mc{X}};\mc{Y})$ and quantum MACs by $\mc{Q}_N(\bm{\mc{X}};\mc{Y})$. We will use $\mc{Q}_N$ and $\mc{C}_N$ to denote general $N$-sender channels with arbitrary input and output alphabets.

Despite the fact that both classical and quantum MACs can be described using Born's rule (i.e. Eqs. \eqref{Eq:classical-Born} and \eqref{Eq:quantum-Born}, respectively), classical MACs admit a much simpler characterization.  The state $\sum_{i=1}^Np_i\op{\mbf{e}_i}{\mbf{e}_i}$ can be understood simply as a classical particle that is sent along path $i$ with probability $p_i$.  A local NPE operation then reduces to probabilistically applying some local channel that either lets the particle continue along its respective path or blocks it from reaching the receiver $\msf{B}$, i.e. either $\mc{E}^{(\text{vac})}$ or the identity map is performed.  With probability $q_i(0|x_i)$ the particle is blocked by party $\msf{A}_i$ for input choice $x_i$, and with probability $q_i(\mbf{e}_i|x_i)$ it is transmitted.  Hence if the input state is $\op{\mbf{e}_i}{\mbf{e}_i}$, then the state received by $\msf{B}$ is 
\begin{align}
    \sigma_{x_i}&=\bigotimes_{j\not=i}\op{0}{0}^{\msf{A}_j}\otimes \mc{E}^{\msf{A}_i}_{x_i}(\op{1}{1})\notag\\
    &= q_i(\mbf{e}_i|x_i)\op{\mbf{e}_i}{\mbf{e}_i}^{\msf{A}_1\cdots \msf{A}_N}+q_i(0|x_i)\op{0}{0}^{\msf{A}_1\cdots \msf{A}_N}.\notag
\end{align}
On the decoding end, party $\msf{B}$ examines each path to see if it contains a particle.  Output $b$ is produced with probability $d(b|\mbf{e}_i)$ when a particle is received along path $i$ and with probability $d(b|0)$ when no particle is received.  Hence the channel obtained after averaging over all input states is 
\begin{align}
\label{Eq:classical-MAC-decomposition}
    p(y|x_1,\cdots,x_N)=\sum_{i=1}^Np_i [&d(y|0)q_i(0|x_i)\notag\\
    +&d(y|\mbf{e}_i)q_i(\mbf{e}_i|x_i)].
\end{align}
The set $\mc{C}_N(\mc{X}_1,\cdots\mc{X}_N;\mc{Y})$ consists of MACs that can be written in this form.

\subsection{Coherence-assisted communication}
\label{Sect:coherence-assisted}

\begin{figure}[t]
\centering
  \includegraphics[width=.45\textwidth]{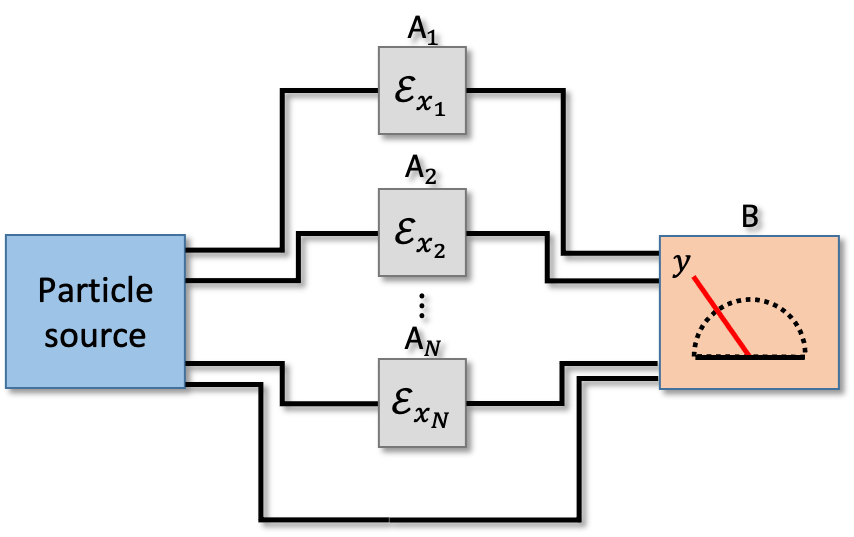}
  \caption{A coherence-assisted protocol allows for an unperturbed side channel through which the particle can traverse coherently to the decoder.}
  \label{fig:coherence-assisted-diagram}
\end{figure}

Thus far we have focused on scenarios where the number of senders equals the number of paths through which the particle source emits the particle.  We can generalize this model by allowing for extra paths that are not acted upon by an sender (Fig. \ref{fig:coherence-assisted-diagram}).  We refer to these as \textit{coherence-assisted protocols},
%We can alternatively ask what happens if the number of senders and the number of paths differ. Obviously, the cases where the number of senders exceeds the number of paths are already included in our framework since the extra senders are not involved in the communication, and thus we can simply ignore them. On the other hand, our previous framework does not include cases where there are more paths than senders. We will call protocols that contain extra paths that are not acted on by any sender \textbf{coherence-assisted protocols} (Fig. \ref{fig:coherence-assisted-diagram}). 
with the extra paths being called called assistance paths.  %Protocols that do not use assistance paths will called \textbf{unassisted protocols}.  
Note that since the assistance paths are not touched by any encoding operation, we can without loss of generality combine amplitudes of multiple assistance paths into one assistance path. %and it will therefore suffice to consider only one assistance path \textcolor{red}{is this explanation clear?}. 
Intuitively, the assistance path can serve as a phase reference for the other paths, which can help the receiver better discriminate the encoded messages. One the other hand, as we will see in Theorem \ref{thm:classical-rate-sum-bound} below, this assistance path cannot enhance communication rate when the source is a classical particle. We let $\mc{Q}^{\ass}_N(\bm{\mc{X}};\mc{Y})$ denote the family of all coherence-assisted channels built by $N$ parties using a single quantum particle and NPE operations. %\todo{(shall we use $\mc{Q}^{\text{ass}}_N$ instead of $\mc{Q}^*_N$, the review paper use the similar notation for entanglement-assisted classical communication)}

An analogy can be drawn here to entanglement-assisted communication \cite{Bennett-1992a, Bennett-2002a}, in which entanglement is shared between the senders and receiver.  In fact, one could imagine in Fig. \ref{fig:coherence-assisted-diagram} that the particle is coherently distributed to the receiver along the assisted path prior to the encoding of the senders. Then the scenario becomes conceptually equivalent to the entanglement-assisted setup except that the shared resource between senders and receivers is coherence in single-particle spatial modes rather than coherence in multi-particle states \cite{vanEnk-2005a}. 

\subsection{Communication rates of MACs}

In this work we consider the achievable communication rates of the classical multiple-access channels as constructed in the previous subsection.  Roughly speaking, a rate tuple $(R_1,\cdots,R_N)$ is achievable for a given MAC if for every $\epsilon>0$ and $n$ sufficiently large, each sender $i$ can send $2^{n R_i}$ possible messages with average error no greater than $\epsilon$ (see Ref. \cite{Cover-2006a} for details).
%Here we recall the definition of an achievable rate tuple of a multiple-access channel.
%\begin{definition}
%For a MAC $p(y|\bm{x})$, a rate tuple $(R_1,\cdots,R_N)$ is called \textbf{achievable} if for every $\epsilon>0$, there exists sufficiently large $n$, encoders $f_i:\lfloor 2^{n R_i}\rfloor\to \mc{X}_i^n$ and decoder $\phi:\mc{Y}^n\to\lfloor 2^{ n R_1}\rfloor \times\cdots\times \lfloor 2^{ n R_N}\rfloor$ such that
%\begin{align}
%\label{Eq:MAC-avg-err}
%   &\frac{1}{2^{n( R_1+\cdots+R_N)}}\sum_{m_1=0}^{\lfloor 2^{ n R_1}\rfloor}\cdots\sum_{m_N=0}^{\lfloor2^{ n R_N}\rfloor} \notag\\
%   &\times p^n(\phi^{-1}(m_1,\cdots,m_N)|f_1(m_1),\cdots, f_N(m_N)) > 1-\epsilon,
%\end{align}
%where $\mc{X}_i$ and $\mc{Y}$ are the input and output alphabets of each individual channel $p(y|\bm{x})$. The \textbf{rate region} for a MAC is then defined as the collection of all achievable rate tuples.
%\end{definition}
%\noindent Equation \eqref{Eq:MAC-avg-err} says that the receiver can decode all $2^{nR_1}\times\cdots\times2^{nR_N}$ possible messages sent by the receivers with average error no greater than $\epsilon$.  
Remarkably the achievable rate region of an $N$-sender MAC has a single-letter characterization in terms of the conditional mutual information, which, for random variables $X_1,X_2,Y$, is defined as $I(X_1:Y|X_2)=I(X_1X_2:Y)-I(X_2:Y)$.
\begin{proposition}[\cite{Liao-1972a, Ahlswede-1973a, Cover-2006a}]
\label{Prop:MAC}
A rate tuple $(R_1,\cdots,R_N)$ for MAC $p(y|\bm{x})$ is achievable if and only if it lies in the closure of the convex hull of all rate tuples satisfying
\begin{equation}
    R_S \leq I(X_S:Y|X_{S^C}), \qquad \forall S\subset\{1,\cdots N\}
\end{equation}
for some product distribution $p(x_1) \cdots p(x_N)$ over the input alphabet $\bm{\mc{X}}$. Here in a slight abuse of notation we denote $X_S\coloneqq \times_{i\in S}X_i$ and $R_S:=\sum_{i\in S}R_i$. In particular, for two parties, the achievable rate region is the convex hull of all rate pairs satisfying
\begin{align}
    R_1&\leq I(X_1:Y|X_2)\notag\\
    R_2&\leq I(X_2:Y|X_1)\notag\\
    R_1+R_2&\leq I(X_1X_2:Y),
\label{eq:rateregion_constrian}
\end{align}
for product distributions $p(x_1)p(x_2)$.
\end{proposition}

For the purpose of this investigation, we will be mainly interested in the largest amount of information that can be jointly sent by the senders.  In our framework, this corresponds to the largest rate-sum $R\coloneqq\sum_{i\in\{1,\cdots,N\}} R_i$ that can be achieved using a MAC constructed from a single particle.

\begin{comment}
we are interested not only in the rate region of one particular MACs, but also the union of all rate regions for classical or quantum MACs. This union of rate regions represent all rate tuples that are achievable classically or quantum mechanically. Additionally, in our work, we will focus heavily on the rate sum $R\coloneqq\sum_{i\in\{1,\cdots,\N\}} R_i$, characterizing the total amount of information that can be jointly sent by the senders. 
\begin{definition}
\todo{Do we need keep the definition for rate region, or just define their convex hull? Since the former one is not discussed too much in our context.} \textcolor{blue}{I think we don't need this definition here.} $R(\mc{C}_N)$ consists of all rate tuples that can be achieved using an MAC in $\mc{C}_N$. We analogously define the \textbf{$N$-sender quantum rate region} $R(\mc{Q}_N)$. The convex hull of these regions will be denoted by $\ol{R}(\mc{C}_N)$ and $\ol{R}(\mc{Q}_N)$, respectively.
\end{definition}
\begin{remark}
It is not difficult to see that $\ol{R}(\mc{C}_N)$ corresponds to the communication rates achievable when shared randomness is distributed to the senders and receiver.
\end{remark}
\end{comment}

\begin{figure*}[t]
\centering
  \includegraphics[width=0.8\textwidth]{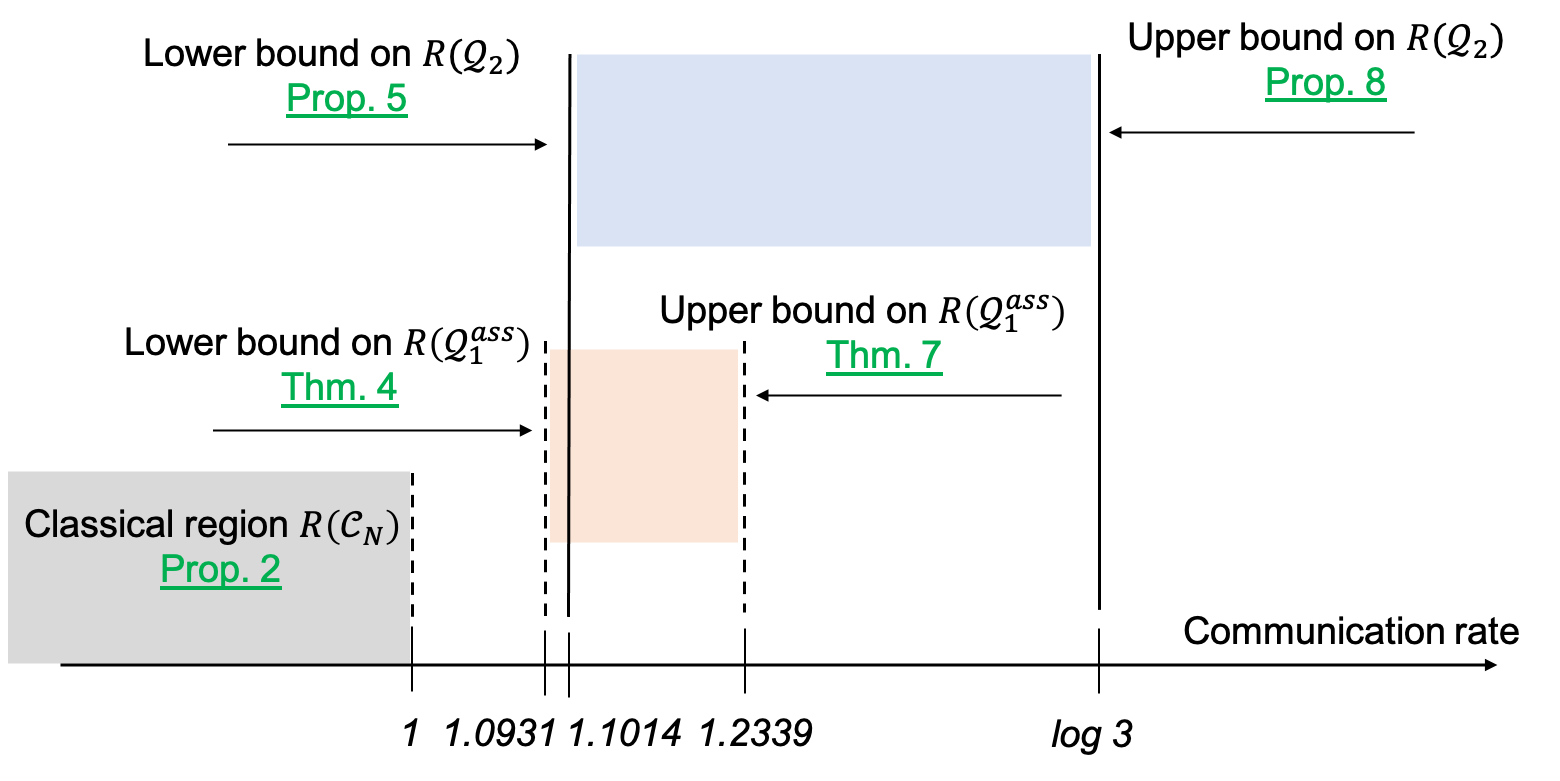}
  \caption{An illustration of our bounds on the one-sender coherence-assisted communication rate $R(\mc{Q}_1^{\ass})$ and two-sender unassisted communication rate sum $R(\mc{Q}_2)$. Each bound is established via the labeled proposition or theorem.}
  \label{fig:communication rate}
\end{figure*}
\subsection{The accessible information and Holevo information}
\label{sec:acc-info-holevo-info}
As described in the previous sections, each communication protocol using a single particle consists of three elements: a choice of the initial one-particle state $\rho$, an encoding strategy which specifies a family of NPE encoding operations $\{\mc{E}_{x_i}^{\msf{A}_i}\}$, and the decoding measurement $\{\Pi_y\}$. We will be interested in optimizing the joint achievable communication rate under this framework, and to do so, we split the full optimization into two parts. Every choice of initial state, encoding strategy, and prior product distribution $p(\bm{x})=p(x_1)\cdots p(x_N)$ over the messages gives rise to the classical-quantum (cq) state
\begin{equation}
    \sigma^{\bm{\msf{XA}}} = \sum_{\bm{x}} p(\bm{x}) \state{\bm{x}}^{\bm{\msf{X}}} \otimes \sigma_{\bm{x}}^{\bm{\msf{A}}} \label{Eq:encoded-cq}
\end{equation}
where $\sigma_{\bm{x}}^{\bm{\msf{A}}}=\mc{E}_{x_1}^{\msf{A}_1}\otimes\cdots\otimes\mc{E}_{x_N}^{\msf{A}_N}(\rho)$.  For each such cq state, when a POVM $\{\Pi_y\}$ is performed on systems $\bm{\msf{A}}$, the resulting joint probability distribution can be described by the classical-classical (cc) state
\begin{equation}
    \sigma^{\bm{\msf{X}}\msf{Y}} = \sum_{\bm{x},y} p(\bm{x})p(y|\bm{x}) \state{\bm{x}}^{\bm{\msf{X}}}\otimes\state{y}^{\msf{Y}}.
\end{equation}
where $p(y|\bm{x})=\tr(\Pi_y\sigma_{\bm{x}})$ is the constructed MAC in $\mc{Q}_N(\bm{\mc{X}};\mc{Y})$.  If $\bm{X}$ denotes the random variable over all $N$ messages and $Y$ denotes the output variable for the receiver, then the information obtained by the receiver about $\bm{X}$ is the mutual information $I(\bm{X}:Y)_{\sigma^{\bm{\msf{X}}\msf{Y}}}$. Optimizing over all POVMs quantifies the so-called accessible information of the cq state $\sigma^{\bm{\msf{XA}}}$,
\begin{equation}
    I_{acc}(\sigma^{\bm{\msf{XA}}}) \coloneqq \max_{\{\Pi_y\}}I(\bm{X}:Y)_{\sigma^{\bm{\msf{X}}\msf{Y}}}.
\end{equation}
We then further optimize the accessible information over all valid cq states (i.e. those having the form of Eq. \eqref{Eq:encoded-cq}),
\begin{align}
    R(\mc{Q}_N)\coloneqq\max_{\sigma^{\bm{\msf{XA}}}}I_{acc}(\sigma^{\bm{\msf{XA}}}).
    \label{eq:optimized-accessible-information}
\end{align}
Thanks to Proposition \ref{Prop:MAC}, $R(\mc{Q}_N)$ captures the largest communication rate-sum that quantum mechanics allows when using a fixed encoding strategy and decoding measurement on each particle.  This is the central quantity of interest in this paper.

Lower bounds of $R(\mc{Q}_N)$ are given by $R(\mc{Q}_N) \geq I_{acc}(\sigma^{\bm{\msf{XA}}}) \geq I(\bm{X}:Y)_{\sigma^{\bm{\msf{X}}\msf{Y}}}$, with $I(\bm{X}:Y)_{\sigma^{\bm{\msf{X}}\msf{Y}}}$ arising from any explicit protocol.  On the other hand, the celebrated Holevo's bound limits the accessible information as
\begin{align}
    I_{acc}(\sigma^{\bm{\msf{XA}}}) \leq \chi(\sigma^{\bm{\msf{XA}}})
    \end{align} 
where $\chi(\sigma^{\bm{\msf{XA}}})\coloneqq I(\bm{X}:\bm{A})_{\sigma^{\bm{\msf{XA}}}} = S\left(\sum_{\bm{x}}p({\bm{x}})\sigma_{\bm{x}}^{\bm{\msf{A}}}\right) - \sum_{\bm{x}}p({\bm{x}})S\left(\sigma_{\bm{x}}^{\bm{\msf{A}}}\right)$ is called the Holevo information \cite{Holevo-1973b}.  Therefore, a natural upper bound for $R(\mc{Q}_N)$ is 
\begin{equation}
    R(\mc{Q}_N) = \max_{\sigma^{\bm{\msf{XA}}}} I_{acc}(\sigma^{\bm{\msf{XA}}}) \leq \chi(\mc{Q}_N) \coloneqq \max_{\sigma^{\bm{\msf{XA}}}} \chi(\sigma^{\bm{\msf{XA}}}).
    \label{eq:optimized-holevo-information}
\end{equation}
%\todo{(It is hard to tell the differences between $I(\bm{X}:\bm{A})_{\sigma^{\bm{\msf{XA}}}}$ and $I(X:Y)_{\sigma^{\bm{\msf{X}}\msf{Y}}}$. Besides, $R_1$ and $R_2$ were used for achievable rate, $R(\mc{Q}_N)$ looks too similar to those quantities)} 
Again, the maximization is over cq state having the form of Eq. \eqref{Eq:encoded-cq}. 

In addition to providing an upper bound, the Holevo information $\chi(\sigma^{\bm{\msf{XA}}})$ admits an operational interpretation \cite{holevo-1998,Schumacher-1997a} within our one-particle communication framework. Namely, it captures the scenario in which the senders prepare independent and identically distributed (i.i.d.) copies of $\sigma^{\bm{\msf{XA}}}$, yet the receiver is allowed to perform joint decoding measurement across all copies. In the asymptotic limit, the largest amount of information that the receiver can gain is exactly the Holevo information $\chi(\sigma^{\bm{\msf{XA}}})$. Therefore, the optimized Holevo information $\chi(\mc{Q}_N)$ represents the ultimate amount of information that can be transmitted by $N$ senders using a fixed single-particle encoding scheme.
%by a single particle assuming the receiver can perform joint detection on i.i.d. copies of encoded particles.  
We similarly let $R(\mc{Q}^{\ass}_N)$ and $\chi(\mc{Q}^{\ass}_N)$ be defined as in Eqs. \eqref{eq:optimized-accessible-information} and \eqref{eq:optimized-holevo-information}, respectively, except with the maximum now taken over all cq states $\sigma^{\bm{\msf{XA}}}$ built using a coherence assistance path.

While $R(\mc{Q}_N)\leq \chi(\mc{Q}_N)$ and $R(\mc{Q}^\ass_N)\leq \chi(\mc{Q}^\ass_N)$, for general $N$ these bounds appear to be quite loose.  For example, we show below that $\chi(\mc{Q}_N)\geq \log N$ and $\chi(\mc{Q}^\ass_N)\geq \log(N+1)$.  On the other hand, the best lower bounds on $R(\mc{Q}_N)$ and $R(\mc{Q}^\ass_N)$ we obtain do not even exceed $1.13$.  While this bound still exceeds the largest classical rate, which is the main focus of this paper, its divergence from the Holveo information reflects the strong communication degradation that arises when restricting to single-copy measurements.

\section{Theoretical Results}\label{sect:theoretical-results}
%\todo{(Add summery of results here? subsection by subsection, with different protocols. For each subsection, introduce each protocol at first with explicit (1) Source state; (2)Encoding op; (3) Decoding op; (4) And quantum enhancement. Then explain, the optimally, or connection of these protocols to the others)}.\par 

Having established our communication model, we now probe the theoretical limits of single particle communication in both the classical and quantum settings.  Our main goal is to place bounds on the communication rates introduced in the previous section.  For simplicity, we focus on multiple-access channels with binary and ternary inputs/outputs.  In Section \ref{subsec: Clasical rate} we compute the ultimate communication rates using a classical particle, which serve as thresholds for our quantum protocols. In Section \ref{subsec: Quantum rate} we construct explicit quantum-enhanced communication protocols. Lower bounds on $R(\mc{Q}^{\text{ass}}_1)$ and $R(\mc{Q}_2)$ are presented in Sections \ref{Sect:one-sender} and \ref{sect:two-sender-MAC}, respectively; for $N\geq 2$ lower bounds on $R(\mc{Q}_N)$ and $R(\mc{Q}^{\text{ass}}_N)$ are provided in Section \ref{Sect:general_N_encoding}; and finally in Section \ref{Sect:Holevo-info} we show that both $\chi(\mc{Q}_N)$ and $\chi(\mc{Q}^{\text{ass}}_N)$ grow as $\log N$.

%Our main goal in this section is to study and compare $R(\mc{Q}_N)$, $R(\mc{Q}_N^*)$, $R(\mc{C}_N)$, and $R(\mc{C}_N^*)$ theoretically. In Sec.~\ref{subsec: Clasical rate}, we upper bound the the achievable rate sums of MACs constructed with a classical particle, which will serve as the benchmark for the quantum-enhanced protocols that we subsequently present. For the quantum case, instead of explicitly calculating $R(\mc{Q}_N)$ and $R(\mc{Q}_N^*)$, we give lower bounds of these quantities by finding explicit protocols. Specifically, we will present a one-sender assisted protocol in Sec.~\ref{Sect:one-sender}, two-sender assisted/unassisted protocols in Sec.~\ref{sect:two-sender}, and the results are generalized to more senders in Sec.~\ref{sect:n-sender}. Besides, we compare the two-sender MACs in detail in terms of the achievable rate region in Sec.~\ref{sect:two-sender}. These achievable lower bounds in quantum MACs provide examples of nontrivial quantum-classical gaps and can be used to show the quantum advantage in the multiple-access communication scenario even in the most basic level of single particles.     

\subsection{Classical MACs}

\label{subsec: Clasical rate}

\subsubsection{$N$-party rate-sum}

We begin by establishing the intuitive upper bound of one bit for the $N$-party rate-sum using a single classical particle. 
The following proposition places a fundamental bound on $N$-party communication within our framework.
\begin{proposition}
    \label{thm:classical-rate-sum-bound}
    $R(\mc{C}_N)=1 \;\forall\,N$. That is, we can communicate at most 1 bit of information using a classical particle.  Furthermore, an assistance path does not help in the classical setting.
\end{proposition}
\begin{proof}
    We first show that $R(\mc{C}_N)\leq 1$. According to Eq.~\eqref{Eq:classical-MAC-decomposition}, any channel in $\mc{C}_N$ admits the decomposition 
    \begin{align}
        p(y|x_1\cdots x_N) = \sum_i p_i \sum_{m=0,\mbf{e}_i} d(y|m)q_i(m|x_i)
    \end{align}
    where $d(y|m)$ and $q(m|x_i)$ are conditional probability distributions associated with the decoder and the encoder, respectively. By convexity of mutual information $I(X_1\cdots X_N:Y)$ with respect to the underlying channel, we can conclude that the rate sum is maximized by channels of the form $p(y|x_1\cdots x_N)=p(y|x_i)=\sum_{m=0,\mbf{e}_i} d(y|m)q_i(m|x_i)$. However, capacities of these channels cannot exceed one bit since $d(y|m)$ is essentially a classical post-processing map, and $q_i(m|x_i)$ is a channel with binary outputs.  

    On the other hand, suppose the initial state is $\state{\mbf{e}_i}$, the $i$-th sender $\msf{A}_i$ encodes information by either annihilating the particle or preserving the particle, and the receiver performs measurement in the particle number basis. In this case, $\msf{A}_i$ can send 1 bit of information, while other senders cannot send any information. So, the total amount of transmitted information is one bit, and therefore $R(\mc{C}_N)\geq1$. To see that an assistance path does not help, observe that $R(\mc{C}_N)=1$ holds for arbitrary $N$, and an assistance path can be seen as a special case of $\mc{C}_{N+1}$ where the $(N+1)$-th party acts trivially.
\end{proof}

\subsubsection{Classical canonical form}

We next turn to the problem of identifying achievable rate tuples using a single classical particle.  This task is simplified by recognizing that every canonical MAC can be obtained from a canonical MAC combined with stochastic encoders and a stochastic decoder.  By the data processing inequality, stochastic post-processing cannot increase the rate region, and the same is true for stochastic pre-processing (Problem 14.5 in \cite{Csiszar-2011a}).  Therefore, if $(R_1,\cdots,R_N)$ is a rate tuple achievable by some single-particle classical MAC, then it is also achievable by a canonical one defined below.
\begin{proposition}
For arbitrary input and output sets $\mc{X}_1\times\cdots\times\mc{X}_N$ and $\mc{Y}$, every MAC in $\mc{C}_N(\mc{X}_1,\cdots\mc{X}_N;\mc{Y})$ can be seen as arising from a canonical MAC that has binary inputs for each sender and $N+1$ outputs for the receiver.
\label{prop:arbitrary input-output classical}
\end{proposition}
\begin{proof}
For a given classical state $\rho= \sum_{k=1}^N p_k\state{\mbf{e}_k}$ and induced MAC $p(y|x_1,\cdots,x_N)$ having the form of Eq. \eqref{Eq:classical-MAC-decomposition}, define the canonical MAC with transition probabilities
\begin{subequations}
\begin{align}
\label{Eq:Canonical-a}
   \wt{p}(k|j_1,\cdots,j_N)&=\begin{cases} p_k\;\;&\text{if $j_k=1$}\\0\;\;&\text{if $j_k=0$}
    \end{cases};\\
    \label{Eq:Canonical-b}
    \wt{p}(0|j_1,\cdots,j_N)&=\sum_{\substack{k\\\text{such that $j_k=0$}}}p_k.
    \end{align}
    \end{subequations}
This channel likewise has the form of Eq. \eqref{Eq:classical-MAC-decomposition} and therefore belongs to $\mc{C}([2],\cdots,[2];[N+1])$.  Also, define local pre-processing stochastic maps $\wt{q}_i:\mc{X}_i\to\{0,1\}$ with $\wt{q}_i(0|x_i)=q_i(0|x_i)$ and $\wt{q}_i(1|x_i)=q_i(\mbf{e}_i|x_i)$, along with a post-processing stochastic map $\wt{d}:\{0,1\,\cdots,N\}\to\mc{Y}$ by $\wt{d}(y|k)=d(y|\mbf{e}_k)$ for $k=1,\cdots, N$ and $\wt{d}(y|0)=d(y|0)$. Then it is straightforward to verify that
\begin{align} 
p&(y|x_1,\cdots,x_N)
=\sum_{k=0}^N\sum_{j_1=0}^1\cdots\sum_{j_N=0}^1\{\wt{d}(y|k)\notag\\
&\;\;\times \wt{p}(k|j_1,\cdots,j_N)\wt{q}_1(j_1|x_1)\cdots \wt{q}_N(j_N|x_N)\}.
\label{Eq:MAC-canonical}  
\end{align}
\end{proof}

\subsubsection{Two-sender classical rate regions}

\begin{figure}[t]
\centering
  \includegraphics[width=.8\linewidth]{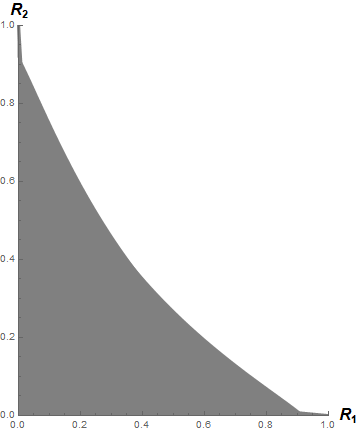}
  \caption{The shaded region is the union of all achievable rate pairs as the weight $\lambda$ of the source state $\rho_\text{cl}=\lambda\state{\mbf{e}_1}+(1-\lambda)\state{\mbf{e}_2}$ varies over interval $[0,1]$.}
  \label{Fig:Rate-region-CC-full}
\end{figure}

We now turn to the rate regions for two-sender communication. 
 Consider the canonical MAC $p(y|\bm{x})$ that is generated by a classical particle $\rho_\text{cl}=\lambda\state{\mbf{e}_1}+(1-\lambda)\state{\mbf{e}_2}$ and having the structure of Eqs. ~\eqref{Eq:Canonical-a} and \eqref{Eq:Canonical-b}.  Since $N=2$, the canonical MAC is characterized by the single parameter $\lambda=p_1$, and the transition probabilities are given by
\begin{align}
    1&=p(00|00)\notag\\
    \lambda&=p(01|01)=p(00|10)=p(01|11)\notag\\
    1-\lambda&=p(10|10)=p(00|01)=p(10|11).
\end{align}
 For a fixed $\lambda\in[0,1]$, and prior $p(x_1)p(x_2)$ the achievable rate pairs $(R_1,R_2)$ are determined by Proposition \ref{Prop:MAC}, which follows a pentangon constrained by Eq.~\ref{eq:rateregion_constrian}. Combining all these regions with with fixed $\lambda\in[0,1]$ but different priors $p(x_1)p(x_2)$, we could obtained the achievable rate region of a specific MAC. \par

We are now interested in computing the union of all achievable rate regions as $\lambda$ is varied within the interval $[0,1]$.  This will yield the total collection of all asymptotic rate pairs $(R_1,R_2)$ feasible by a MAC built using a single classical particle.\par

%It should be emphasized that the non-convexity in Fig. \ref{Fig:Rate-region-CC-full} follows from the non-convexity of the parametric curves $(R^*_1(\lambda),R^*_2(\lambda))$ and $(R^{**}_1(\lambda),R^{**}_2(\lambda))$.  
Note that a rate pair $(R_1,R_2)$ lies in the enclosed region of Fig. \ref{Fig:Rate-region-CC-full} if and only if it is achievable using many copies of the \textit{same} source state $\rho_\text{cl}=\lambda\state{\mbf{e}_1}+(1-\lambda)\state{\mbf{e}_2}$, and the union of these rate pairs evidently forms a non-convex set.  However, if we relax this i.i.d. constraint and allow $\lambda$ to vary across the multiple uses, then more rate pairs are accessible by time-sharing.  In this case, the collection of achievable rate pairs is just the convex hull of the region in Fig. \ref{Fig:Rate-region-CC-full}, i.e. a triangle with outer vertices $(1,0)$ and $(0,1)$.  

\subsection{Quantum MACs}

\label{subsec: Quantum rate}

\subsubsection{Surpassing the classical bound with one sender and coherence assistance}
\label{Sect:one-sender}
\begin{figure}[b]
    \centering
    \includegraphics[width=.48\textwidth]{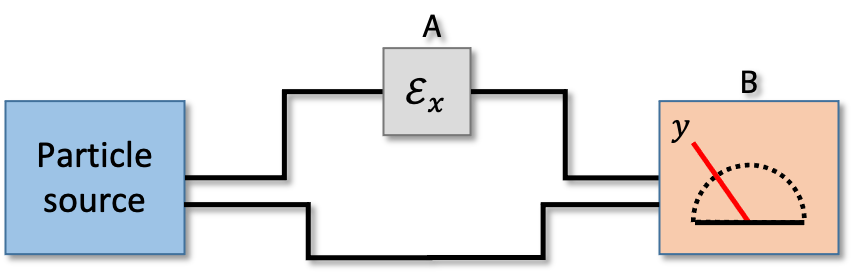}
    \caption{Coherence-assisted communication with one sender.}
    \label{fig:coh-assisted-one-party}
\end{figure}

Given the classical communication bounds established in the previous section, it is natural to consider whether quantum mechanics can do better. We begin by considering the special case of just one sender, and the encoding scheme presented here will generalize as more parties are added.  In the one-sender scenario, if no coherence assistance is used then the whole communication system is simply a two-dimensional space spanned by $\{\ket{0},\ket{1}\}$.  By Holevo's theorem, the communication rate is bounded above by $\log 2 = 1$, and therefore, quantum mechanics offers no advantage over classical physics. However, by leveraging coherence assistance in the sense of Fig. \ref{fig:coh-assisted-one-party}, it is possible to communicate more than one bit of information in the point-to-point scenario. 

To achieve a greater capacity using a single particle, we construct a channel with ternary input symbols. Suppose that the initial state distributed from the particle source is $\ket{\psi}^{\msf{A}\msf{R}}=\cos{\theta}\ket{\mbf{e}_1}+\sin{\theta}\ket{\mbf{e}_2}$ with $\theta\in[0,\pi/2]$.  Note that this describes the most general one-particle state since %\todo{we have assumed that the receiver shares phase reference with the particle source, and 
any relative phase can be absorbed into the definition of $\ket{\mbf{e}_1}$, which we assume is known to the receiver. %Note that we can assume without loss of generality that there is no relative phase factor $e^{i\phi}$ since this relative phase can be absorbed into the POVM at the receiver's end. We can also assume that $\theta\in[0,\pi/2]$ instead of $[0,\pi)$ because $\cos(\frac{\pi}{2}+\theta)=-\cos(\frac{\pi}{2}-\theta)$ and $\sin(\frac{\pi}{2}+\theta)=\sin(\frac{\pi}{2}-\theta)$, and the relative phase change can be dropped. 
For message $x\in\{0,1,2\}$, let the sender $\msf{A}$ encode the state $\ket{\psi}$ according to the following NPE operations:
\begin{equation}
    \begin{cases}
        \mc{E}_0(\rho) = \mc{E}^{\text{(vac)}}(\rho)=\tr(\rho) \ket{0}\bra{0};  \\
        \mc{E}_1(\rho) = \rho;  \\
        \mc{E}_2(\rho) =\mc{E}^{(\alpha)}(\rho)= e^{-i\alpha Z/2}\rho e^{i\alpha Z/2}. 
    \end{cases}
    \label{Eq:one-sender-encoding}
\end{equation}
%Letting $\sigma_x=\mc{E}^{A}_x\otimes\text{id}^R(\op{\psi}{\psi})$, we will show in Sect. \ref{sect:upper-bounds}, that the Holevo information of the cq state $\sigma^{\msf{XAR}} = \sum_x p(x)\state{x}\otimes\sigma_x$ is maximized by the encoding of Eq. \eqref{Eq:one-sender-encoding} along with a message prior of the form $p(0)=1-q$ and $p(1)=p(2)=q/2$.  This motivates us to conjecture that it might also be optimal when using a fixed decoding measurement on each particle.  
Let $\sigma_x=\mc{E}^{\msf{A}}_x\otimes\text{id}^\msf{R}(\op{\psi}{\psi})$ and $\sigma^{\msf{XAR}}=\sum_xp(x)\op{x}{x}\otimes\sigma_x$ be an encoded cq state with the prior distribution over messages have the form $p(0)=1-q$ and $p(1)=p(2)=q/2$.
As shown in Section \ref{Sect:Holevo-info}, the Holevo information $\chi(\mc{Q}^{\ass}_1)$ is attained by this type of cq state.  Hence, we are motivated to conjecture that the encoding scheme of Eq. \eqref{Eq:one-sender-encoding} is also optimal for the single-particle rate $R(\mc{Q}^{\ass}_1)$.  Even if this conjecture fails to be true, the accessible information of $\sigma^{\msf{XAR}}$ still provides a lower bound on $R(\mc{Q}^{\ass}_1)$.

In general, calculating the accessible information of an arbitrary cq state is mathematically challenging. However, in our case, the encoded cq state enjoys the following symmetries: (i) each $\sigma_x$ is block-diagonal in the particle number basis, and  (ii) $q/2\cdot \sigma_1$ and $q/2\cdot \sigma_2$ are related by a reflection across the line $y=x\tan(\alpha/2)$ in the $x-y$ plane of the Bloch sphere.  Using similar arguments to those in Ref. \cite{Frey-2006}, we find (see the Supplemental Material) that $\alpha=\pi$ provides an optimal encoding.  Further analysis then shows that the accessible information is maximized by a prior probability $q$ and coherence angle $\theta$ in the source state that together satisfy a pair of transcendental equations.  Solving these equations numerically leads to the following theorem.

%is symmetric under $\ket{00}\leftrightarrow-\ket{00}$ and $\ket{\mbf{e}_1}\leftrightarrow e^{i\alpha}\ket{\mbf{e}_1}$. These symmetries considerably simplify the problem. Using these symmetries, we were first able to prove that the optimal $\alpha=\pi$. Moreover, setting $\alpha=\pi$, we can analytically calculate the accessible information of $\sigma^{\msf{XAR}}$ for any $q$ and $\theta$. By taking derivatives with respect to $q$ and $\theta$, we arrive at two transcendental equations that the optimal $q$ and $\theta$ must satisfy. Solving these equations numerically, we obtain the optimal $q$ and $\theta$ and the corresponding accessible information. Our result is summarized by the theorem below.

\begin{theorem}\label{thm:one-sender-acc-info}
There exists a one-sender coherence-assisted communication protocol that sends approximately 1.0931 bits per channel use, i.e., $R(\mc{Q}_1^\ass)\geq 1.0931$. The optimal $(q,\theta)$ that achieves this are approximately $(0.8701,\arccos(\sqrt{0.4715}))$, and the optimal measurement projects into the basis $\{\ket{00},\frac{1}{\sqrt{2}}(\ket{\mbf{e}_1}\pm\ket{\mbf{e}_2})\}$.
\end{theorem}

\noindent Note that the largest accessible information is not attained using a state with uniform superposition across both paths.  Yet, the optimal decoding measurement is a projection into uniform superposition states $\frac{1}{\sqrt{2}}(\ket{\mbf{e}_1}\pm\ket{\mbf{e}_2})$.  When using a source state with uniform superposition across both paths (i.e. $\theta=\pi/4$), the largest communication rate is computed to be $1.0875$.

\subsubsection{Two-sender MACs}
\label{sect:two-sender-MAC}
%\begin{figure}[b]
%    \centering
%    \subfloat[]{\includegraphics[width=0.45\textwidth]{two-sender-non-coh.png}\label{fig:non-coh-assisted-two-senders}}\qquad
%    \subfloat[]{\includegraphics[width=0.45\textwidth]{two-sender-coh-assisted.png}\label{fig:coh-assisted-two-senders}}
    %\caption{The schematics of (a) a two-sender unassisted protocol and (b) a two-sender coherence-assisted protocol.}
%\end{figure}
Let us now add a second sender to the communication picture.  We first consider the scenario of two senders with no coherence assistance.  Thus there are only two paths connecting the source to the receiver, and we borrow ideas from the previous one-sender coherence-assisted protocol, which also has two paths. Let senders $\msf{A}_1$ and $\msf{A}_2$ share the state $\ket{\psi}^{\msf{A}_1\msf{A}_2}=\cos\theta\ket{\mbf{e}_1}+\sin\theta\ket{\mbf{e}_2}$.  Consider first the following binary encoding strategy:
\begin{align}
    \label{eq:two-sender-2,2-input-encoding}
    &\begin{cases}
        \mc{E}^{\msf{A}_1}_{0}(\rho) = \tr(\rho)\state{0};\\
        \mc{E}^{\msf{A}_1}_{1}(\rho) = \rho;
        \end{cases}\notag\\
        &\begin{cases}
         \mc{E}^{\msf{A}_2}_{0}(\rho)=\rho; \\
        \mc{E}^{\msf{A}_2}_{1}(\rho) = e^{-i\alpha Z/2}\rho e^{i\alpha Z/2}.
        \end{cases}
\end{align}
Observe that $\mc{E}_0^{\msf{A}_1}\otimes\mc{E}_0^{\msf{A}_2}(\op{\psi}{\psi})=\mc{E}_0^{\msf{A}_1}\otimes\mc{E}_1^{\msf{A}_2}(\op{\psi}{\psi})$, and so there only three distinct encoded states.  In fact, if $\msf{A}_1$ has prior probabilities $\{1-q,q\}$ over messages $\{0,1\}$ and $\msf{A}_2$ has uniform prior probabilities over the messages,  then the resulting cq state $\sigma^{\msf{X}\msf{A}_1\msf{A}_2}$ is equivalent to the cq state $\sigma^{\msf{X}\msf{A}\msf{R}}$ constructed in the one-sender assisted protocol.  Therefore, by Theorem \ref{thm:one-sender-acc-info}, $\alpha=\pi$ is optimal, and the maximal rate sum achievable with this protocol is 1.0931, which is achieved by the source state $\sqrt{0.4715}\ket{\mbf{e}_1}+\sqrt{0.5285}\ket{\mbf{e}_2}$ and encoding probability $q\approx 0.8701$. 

%\begin{remark}
%As before, the maximal rate sum achievable by the equal superposition state is 1.0875, and the induced channel in this case is
%\begin{align}
%    &p(1|10)=1,\quad p(2|11)=1\notag\\
%    &p(0|00)=\frac{1}{2},\quad p(1|00)=\frac{1}{4},\quad p(2|00)=\frac{1}{4}\notag\\
%    &p(0|01)=\frac{1}{2},\quad p(1|01)=\frac{1}{4},\quad p(2|01)=\frac{1}{4},
%    \label{eq:two-sender-unassisted-equal-superposition-channel}
%\end{align}
%where we identified measurement results $\frac{1}{\sqrt{2}}(\ket{\mbf{e}_1}+\ket{\mbf{e}_2})$, $\frac{1}{\sqrt{2}}(\ket{\mbf{e}_1}-\ket{\mbf{e}_2})$, and $\ket{00}$ with 0, 1, and 2, respectively.
%\end{remark}

\begin{figure}[b]
    \centering
    \includegraphics[width=.5\textwidth]{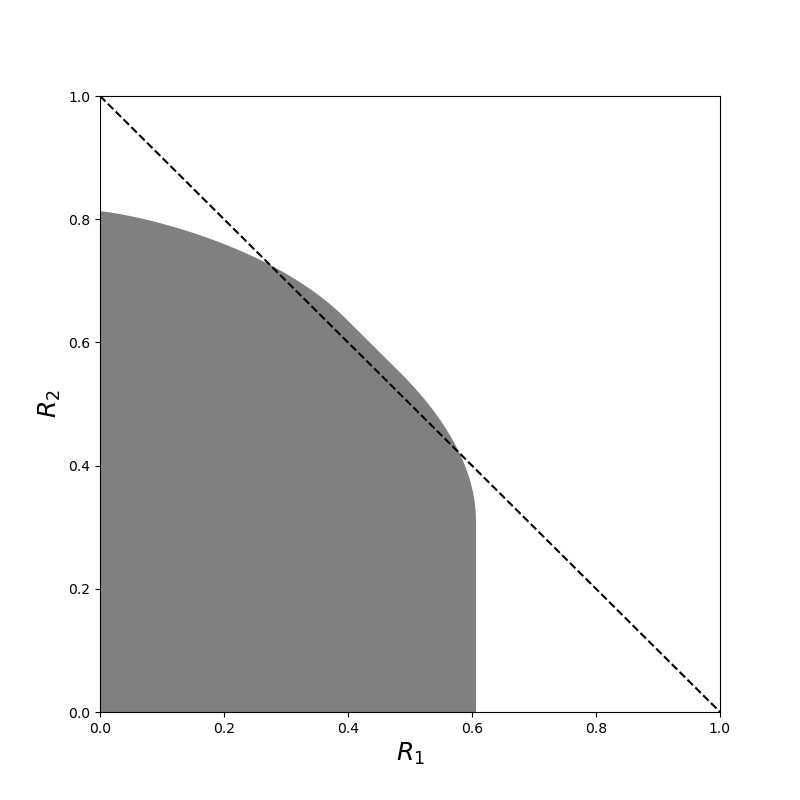}
    \caption{An example of rate region that is achievable by using initial state $\ket{\psi}=\sqrt{1/3}\ket{\mbf{e}_1}+\sqrt{2/3}\ket{\mbf{e}_2}$ and the binary-input protocol.}
    \label{Fig:2-sender-quantum-region}
\end{figure}

\begin{figure}[t]
    \centering
    \includegraphics[width=.5\textwidth]{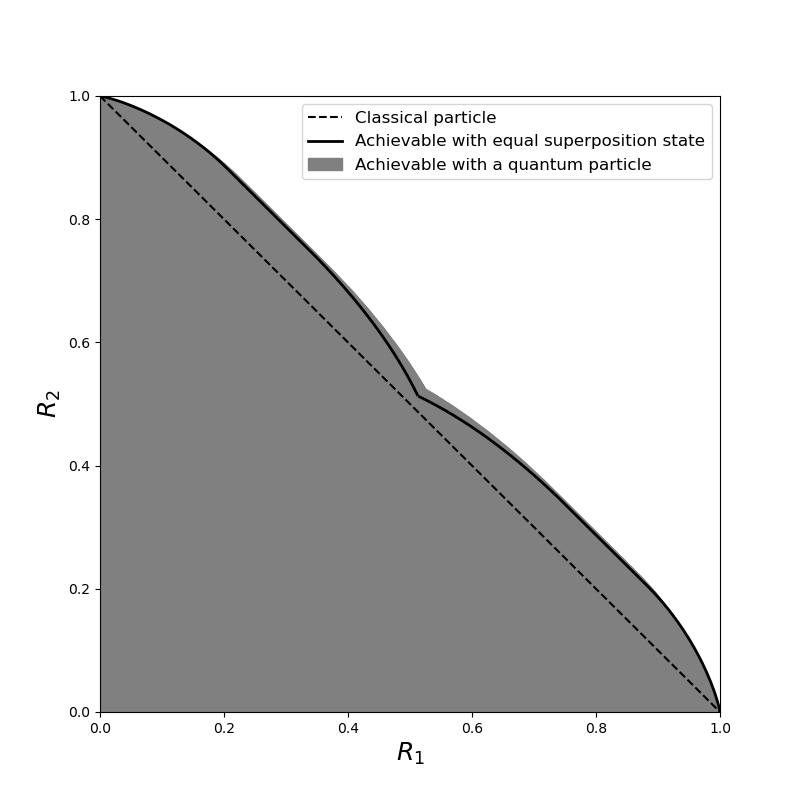}
    \caption{The union of all achievable rate regions using our binary-input protocol (gray area). The solid line represents the boundary of rate region that is achievable using an equal superposition state $\sqrt{1/2}(\ket{\mbf{e}_1}+\ket{\mbf{e}_2})$. The dashed line represent the convex hull of all rate pairs achievable using a classical particle.}
    \label{Fig:2-sender-full-quantum-region}
\end{figure}

The full rate region can also be computed.  For each fixed $\theta\in[0,\pi/2]$, the initial state $\ket{\psi}^{\msf{A}_1\msf{A}_2}=\cos\theta\ket{\mbf{e}_1}+\sin\theta\ket{\mbf{e}_2}$ induces a classical MAC $[2]\times[2]\to[3]$ when using the encoding of Eq. \eqref{eq:two-sender-2,2-input-encoding} and the decoding measurement which projects into the basis $\{\ket{00},\frac{1}{\sqrt{2}}(\ket{\mbf{e}_1}\pm\ket{\mbf{e}_2})\}$.  The specific transition probabilities are found to be
\begin{align}
    &p(0|00)=\cos^2\theta,\quad p(1|00)=p(2|00)=\frac{\sin^2\theta}{2}\notag\\
    &p(0|01)=\cos^2\theta,\quad p(1|01)=p(2|01)=\frac{\sin^2\theta}{2}\notag\\
    &p(1|10)=\frac{1}{2}+\cos\theta\sin\theta,\quad p(2|10)=\frac{1}{2}-\cos\theta\sin\theta\notag\\
    &p(1|11)=\frac{1}{2}-\cos\theta\sin\theta,\quad p(2|11)=\frac{1}{2}+\cos\theta\sin\theta.\notag
\end{align}
The rate region $(R_1,R_2)$ is then found using Proposition \ref{Prop:MAC} (see Fig. \ref{Fig:2-sender-quantum-region}).  As we sweep $\theta$ over the interval $[0,\pi/2]$, the union of all achievable rate pairs using encoding scheme \eqref{eq:two-sender-2,2-input-encoding} is identified in Fig. \ref{Fig:2-sender-full-quantum-region}.  The solid line in this figure indicates the outer boundary on achievable rates using a uniform superposition input state $\ket{\psi}=\frac{1}{\sqrt{2}}(\ket{\mbf{e}_1}+\ket{\mbf{e}_2})$.  These values are noteworthy since they are what we try to experimentally replicate in Section \ref{Sect:experiment}.

We can enhance the rate sum even further if we allow one of the parties to have three inputs. Suppose now that $\msf{A}_2$ encodes with the same ternary operation as in Eq.~\eqref{Eq:one-sender-encoding}, and $\msf{A}_1$ again uses the on-off keying encoding:
\begin{align}
    &\begin{cases}
        \mc{E}^{\msf{A}_1}_{0}(\rho) = \tr(\rho) \state{0};\\
        \mc{E}^{\msf{A}_1}_{1}(\rho) = \rho; 
    \end{cases}\notag\\
    &\begin{cases}
        \mc{E}_0^{\msf{A}_2}(\rho) = \tr(\rho)\ket{0}\bra{0};  \\
        \mc{E}_1^{\msf{A}_2}(\rho) = \rho; \\
        \mc{E}_2^{\msf{A}_2}(\rho) = e^{-i\alpha Z/2}\rho e^{i\alpha Z/2}.
    \end{cases}.
    \label{eq:two-sender-2,3-input-encoding}
\end{align}
Suppose that Alice and Bob's prior probability of message $0$ is $q$ and $q'$, respectively.  Then using the same method of calculating the accessible information of symmetric ensembles (see the Supplemental Material), we again find that the optimal phase encoding is $\alpha=\pi$. This allows us to calculate the accessible information of the encoded cq state for any $q$, $q'$, and $\theta$, which we then maximize.
\begin{proposition}
    There exists a two-sender unassisted communication protocol $[2]\times [3]\to[3]$ that sends 1.1014 bits of information per channel use, i.e., $R(\mc{Q}_2)\geq 1.1014$. The optimal $(q,q',\theta)$ that achieves this are approximately $(0.9197,0.9197,\pi/4)$, and the optimal measurement is given by projecting on the basis $\{\ket{00},\frac{1}{\sqrt{2}}(\ket{\mbf{e}_1}\pm\ket{\mbf{e}_2})\}$.
    \label{prop:two-sender-rate-sum}
\end{proposition}
\noindent Note that unlike in this case of binary encoding, the optimal source state is a uniform superposition across both paths (i.e. $\theta=\pi/4)$.

\subsubsection{A general encoding method for $N\geq 2$ parties without blocking}

\label{Sect:general_N_encoding}

One drawback of the encoding schemes presented in Eqs. \eqref{eq:two-sender-2,2-input-encoding} and \eqref{eq:two-sender-2,3-input-encoding} is that it requires one of the parties to perform an on-off keying (i.e ``blocking'') operation.  While intuitively simple, a reliable implementation of this encoding in an optical setup can be quite demanding.  Here we show that through the use of a coherence assistance path, a rate sum strictly larger than one is always achievable using simple $0,\pi$ phase encoding. The latter means that the sender either acts trivially on the particle or applies a rotation $\mc{E}^{(\pi)}(\rho)=Z \rho Z$.

\begin{figure}[b]
    \centering
    \includegraphics[width=0.45\textwidth]{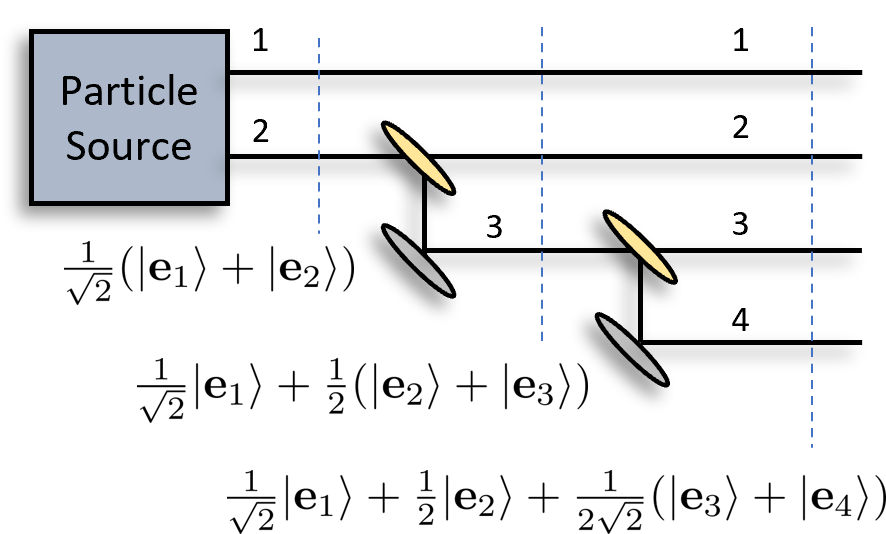}
    \caption{A multi-path state with attenuated amplitudes is generated by a successive application of beam splitters.}
    \label{Fig:multiple_bs}
\end{figure}

\begin{figure}[t]
    \centering
    \includegraphics[width=0.5\textwidth]{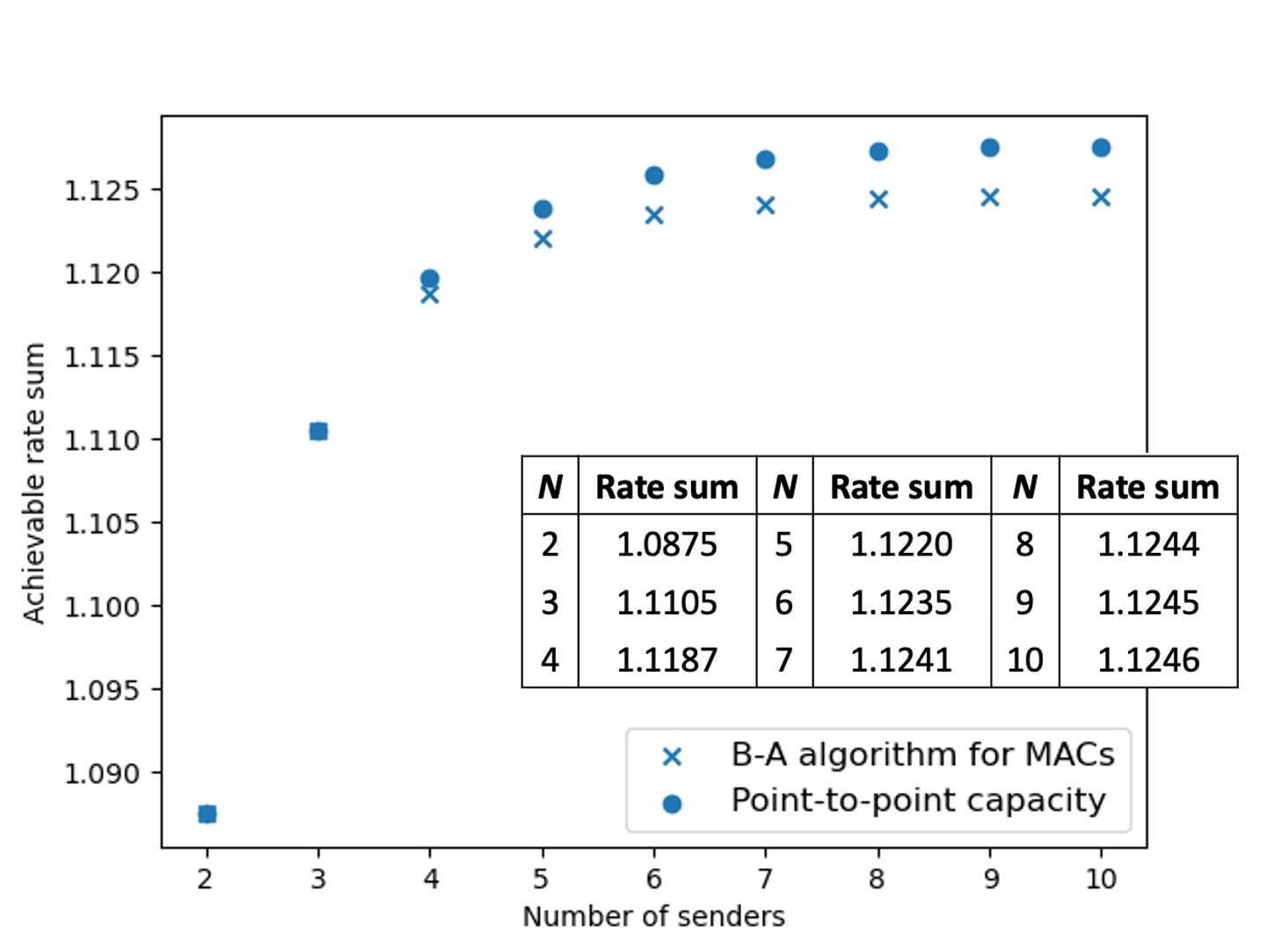}
    \caption{Numerical calculation of rate sums achievable with our $N$-sender protocol [Eq.~\eqref{eq:n-sender-protocol}] for $N$ up to 10.}
    \label{fig:ratesum-vs-N}
\end{figure}

Our protocol involves the idea of creating more paths by successive uses of a beam splitter (see Fig. \ref{Fig:multiple_bs}).  Suppose that at the layer we start with the uniform superposition state $\frac{1}{\sqrt{2}}(\ket{\mbf{e}_1}+\ket{\mbf{e}_2})$.  A beam splitter is inserted along the second path yielding the state $\frac{1}{\sqrt{2}}\ket{\mbf{e}_1}+\frac{1}{2}(\ket{\mbf{e}_2}+\ket{\mbf{e}_3})$.  This is repeated repeatedly until the initial state $\ket{\psi}^{\msf{A}_1\cdots\msf{A}_N\msf{R}} = \sum_{i=1}^{N} \frac{1}{\sqrt{2^i}}\ket{\mbf{e}_i} + \frac{1}{\sqrt{2^{N}}}\ket{\mbf{e}_{N+1}}$ is prepared for $N$ senders $\msf{A}_1,\cdots,\msf{A}_N$ and a coherence assistance path $\msf{R}$. Each sender encodes by applying a $\pi$ phase shift 
\begin{align}
    \mc{E}_{x_i}^{\msf{A}_i}(\rho) = Z^{x_i}\rho Z^{x_i}
    \label{eq:n-sender-protocol}
\end{align}
for message $x_i\in\{0,1\}$ with prior probability $p(x_i)$. Upon receiving the encoded particle, the receiver decodes using the 
projective measurement $\left\{\ket{b_i}\bra{b_i}:
\;i\in[N+1]\right\}$ with orthonormal vectors
\begin{align}
    \ket{b_0}&=\frac{1}{\sqrt{2}}\ket{\mbf{e}_1} + \sum_{i=2}^{N} \frac{1}{\sqrt{2^i}}\ket{\mbf{e}_i} + \frac{1}{\sqrt{2^{N}}}\ket{\mbf{e}_{N+1}} \notag\\
    \ket{b_1}&=-\frac{1}{\sqrt{2}}\ket{\mbf{e}_1} + \sum_{i=2}^{N} \frac{1}{\sqrt{2^i}}\ket{\mbf{e}_i} + \frac{1}{\sqrt{2^{N}}}\ket{\mbf{e}_{N+1}} \notag\\
    \ket{b_2}&=-\frac{1}{\sqrt{2}}\ket{\mbf{e}_2}
    +\sum_{i=3}^{N}\frac{1}{\sqrt{2^{i-1}}}\ket{\mbf{e}_i} + \frac{1}{\sqrt{2^{N-1}}}\ket{\mbf{e}_{N+1}} \notag\\
    &\vdots \notag\\
    \ket{b_N}&=-\frac{1}{\sqrt{2}}\ket{\mbf{e}_N}
    +\frac{1}{\sqrt{2}}\ket{\mbf{e}_{N+1}}. \notag
\end{align}
This induces a classical channel $p(y|x_1,\cdots,x_N)$, and for small $N$ we can numerically compute their capacities using the generalized Blahut-Arimoto algorithm adapted for MACs \cite{blahut-1972,arimoto-1972, Rezaeian-2004}. The result is presented in Fig. \ref{fig:ratesum-vs-N}.
\begin{comment}
\begin{table}[]
    \centering
    \begin{tabular}{c c c c c}
    $N$ & \hspace{6pt}rate sum\hspace{6pt} & \hspace{24pt} & $N$ & \hspace{6pt}rate sum\hspace{6pt} \\\hline\hline
    3 & 1.1105 & & 7 & 1.1241 \\
    4 & 1.1187 & & 8 & 1.1244 \\
    5 & 1.1220 & & 9 & 1.1245 \\
    6 & 1.1235 & & 10 & 1.1246
    \end{tabular}
    \caption{Caption}
    \label{tab:my_label}
\end{table}
\end{comment}
Note that the generalized Blahut-Arimoto algorithm is not guaranteed to converge to the optimal rate sum \cite{Buhler-2011}. However, let us consider $p(y|x_1,\cdots,x_N)$ as a single-sender-single-receiver channel. The original Blahut-Arimoto algorithm does in fact converge to its optimal point-to-point capacity. This point-to-point capacity serves as an upper bound for the rate sum of $p(y|x_1,\cdots,x_N)$ since we are giving senders more power to coordinate. 

For $N=2$, the problem allows for an analytic solution, and we summarize the result in the proposition below.
\begin{proposition}
    There exists a two-sender coherence-assisted communication protocol that \textit{does not require blocking operation or vacuum detection} and sends $\log(17/8)\approx1.0875$ bits per channel use, i.e., $R(\mc{Q}_2^{\text{ass}})\geq1.0875$. The optimal prior probability that achieves this is $p(x_1=0)=1/2$ and $p(x_2=0)=15/17$.
    \label{prop:two-sender-assisted-protocol}
\end{proposition}
As $N$ increases, we numerically find that the rate sum does not increase significantly.  On the one hand, this is not surprising since our encoding strategy uses an initial state $\ket{\psi}$ that places smaller and smaller weight on the paths of additional parties.  However, on the other hand, we have not been able to find any superior coding method, and in fact, many coding schemes (such as the ``fingerprinting'' protocol \cite{Horvat-2021a, Zhang-2022}) have a rate sum that vanishes as $N$ grows large.  A significant open problem is to find upper bounds on the largest $N$-party rate sum using a single quantum particle, which we conjecture will not be too far from the lower bound depicted in Fig. \ref{fig:ratesum-vs-N}.  

The coherence-assisted protocol just described uses only phase encodings. However, it can easily be converted to a coherence-unassisted communication protocol at the expense of needing blocking operations.  To see the idea, consider the case of $N=2$.  In the unassisted protocol, the encoded states $\sigma_{x_1x_2}^{\msf{A}_1\msf{A}_2\msf{R}}=\op{\psi_{x_1x_2}}{\psi_{x_1x2}}$ have the form
\begin{align}
    \ket{\psi_{00}} &= \frac{1}{\sqrt{2}}\ket{\mbf{e}_1}+\frac{1}{2}\ket{\mbf{e}_2}+\frac{1}{2}\ket{\mbf{e}_3} \notag\\
    \ket{\psi_{01}} &= \frac{1}{\sqrt{2}}\ket{\mbf{e}_1}-\frac{1}{2}\ket{\mbf{e}_2}+\frac{1}{2}\ket{\mbf{e}_3} \notag\\
    \ket{\psi_{10}} &= -\frac{1}{\sqrt{2}}\ket{\mbf{e}_1}+\frac{1}{2}\ket{\mbf{e}_2}+\frac{1}{2}\ket{\mbf{e}_3}\notag\\
    \ket{\psi_{11}} &= -\frac{1}{\sqrt{2}}\ket{\mbf{e}_1}-\frac{1}{2}\ket{\mbf{e}_2}+\frac{1}{2}\ket{\mbf{e}_3}.\notag
    \end{align}
Observe that these are made equivalent to the states 
\begin{align}
  \ket{\psi'_{00}} &= \frac{1}{\sqrt{2}}\ket{\mbf{e}_1}+\frac{1}{\sqrt{2}}\ket{\mbf{e}_2}\notag\\
    \ket{\psi'_{01}} &= \frac{1}{\sqrt{2}}\ket{\mbf{e}_1}+\frac{1}{\sqrt{2}}\ket{\mbf{e}_3}\notag\\
    \ket{\psi'_{10}} &= -\frac{1}{\sqrt{2}}\ket{\mbf{e}_1}+\frac{1}{\sqrt{2}}\ket{\mbf{e}_2}\notag\\
    \ket{\psi'_{11}} &= -\frac{1}{\sqrt{2}}\ket{\mbf{e}_1}+\frac{1}{\sqrt{2}}\ket{\mbf{e}_3}\label{Eq:equiv-encoding}
    \end{align}
by a unitary operator that also transforms the measurement vectors into
\begin{align}
    \ket{b'_0} &= \frac{1}{\sqrt{2}}\ket{\mbf{e}_1}+\frac{1}{\sqrt{2}}\ket{\mbf{e}_2}\notag\\
    \ket{b'_1} &= -\frac{1}{\sqrt{2}}\ket{\mbf{e}_1}+\frac{1}{\sqrt{2}}\ket{\mbf{e}_2}\notag\\
    \ket{b'_2} &= \ket{\mbf{e}_3}.\label{Eq:equiv-decoding}
\end{align}
Hence, the states of Eq. \eqref{Eq:equiv-encoding} and measurement of Eq. \eqref{Eq:equiv-decoding} will generate the same transition probabilities as the original MAC.  But since the $\ket{b_i'}$ have no coherence between the $\{\ket{\mbf{e}_1},\ket{\mbf{e}_2}\}$ and $\{\ket{\mbf{e}_3}\}$ subspaces, we can first dephase the $\ket{\psi'_{x_1x_2}}$ across these subspaces without altering the transition probabilities.  Doing so and relabeling $\ket{0}\equiv \ket{\mbf{e}_3}$ leads to states $\sigma_{x_1x_2}$ obtained by the unassisted encoding of Eq. \eqref{eq:two-sender-2,2-input-encoding} (up to a swap $\msf{A}_1\leftrightarrow\msf{A}_2$).  This method of converting a coherence-assisted protocol to an unassisted-protocol generalizes for any $N\geq 2$.

\subsubsection{The single-particle Holevo capacities}

\label{Sect:Holevo-info}

All of the communication rates computed thus far assumes the receiver performs the same measurement on each received quantum particle so as to generate multiple uses of the same classical channel $p(y|\bm{x})$.  While this leads to a definite communication advantage compared to the use of a classical particle, Fig. \ref{fig:ratesum-vs-N} suggests that this advantage is not that dramatic.  On the other hand, if we enlarge the measurement capabilities of the decoder and allow for collective measurements across multiple particle transmissions, then the capacity can be enlarged significantly.  This quantity is the single-particle Holevo information $\chi(\mc{Q}_N)$ as defined in Eq.~\eqref{eq:optimized-holevo-information}, with $\chi(\mc{Q}^{\text{ass}}_N)$ denoting its coherence-assisted form.

Our first result is the calculation of $\chi(\mc{Q}^{\text{ass}}_N)$ for $N=1$.
\begin{theorem}\label{thm:one-sender-holevo}
   \begin{align}\chi(\mc{Q}_1^{\text{ass}}) &= \max_{q,\cos^2\theta\in[0,1]} q h_2(\cos^2\theta)+\cos^2\theta h_2(q) \notag\\
   &= \max_{x\in[0,1]} 2xh_2(x) \notag\\
   &\approx 1.2339,\notag
   \end{align}
   where $h(x)=- x\log_2 x-(1-x)\log_2(1-x)$ is the binary entropy.
\end{theorem}
\noindent Note that since $R(\mc{Q}_1^{\text{ass}}) \leq \chi(\mc{Q}_1^{\text{ass}})$, this shows that the encoding scheme of Theorem \ref{thm:one-sender-acc-info} is not too far from optimal.  The proof of this Theorem \ref{thm:one-sender-holevo} is provided in the Supplemental Material. 
As an intermediate step in our proof, we show that the encoding strategy of Eq. \eqref{Eq:one-sender-encoding} maximizes the Holevo information for each choice of initial state.  Then optimizing over the initial state, we find the maximum in Theorem \ref{thm:one-sender-holevo} is obtained by the values $(q,\cos^2\theta)\approx (0.7035,0.7035)$.

Turning to the $N$-sender case, we find that the single-particle Holevo information grows unbounded, in sharp contrast to the optimized accessible information, which we do not know exactly, but seems to remain bounded for all $N$ despite our best efforts in searching for better protocols.
\begin{proposition}
$\log N\leq\chi(\mc{Q}_N)\leq\log(N+1)$ and $\log(N+1)\leq\chi(\mc{Q}_N^{\text{ass}})\leq\log(N+2)$.
\label{prop: Holevo informtion}
\end{proposition}
\noindent To achieve the lower bounds, the parties use an equal superposition state $\ket{\psi}=\sum_i \frac{1}{\sqrt{N}}\ket{\mbf{e}_i}$ and $0,\pi$ phase encoding.  If each local message $x_i$ has uniform prior over $\{0,1\}$ then, the average encoded state is $\sum_{\bm{x}}p({\bm{x}})\sigma_{\bm{x}}^{\bm{\msf{A}}}=\frac{1}{N}\sum_{i=1}^N\op{\mbf{e}_i}{\mbf{e}_i}$.  Hence, $\chi(\mc{Q}_N)\geq\log N$.  For the assisted case, a similar construction yields $\chi(\mc{Q}_N^{\text{ass}})\geq\log(N+1)$. The upper bounds are simply dimensionality bounds based on the total number of dimensions of the communication system. The lower bound given here is in general not tight. For instance, when $N=2$, Proposition \ref{prop:two-sender-rate-sum} shows that $\chi(\mc{Q}_N)\geq1.1014$.
\begin{figure*}[t]
    \centering
    \includegraphics[width=0.8\textwidth]{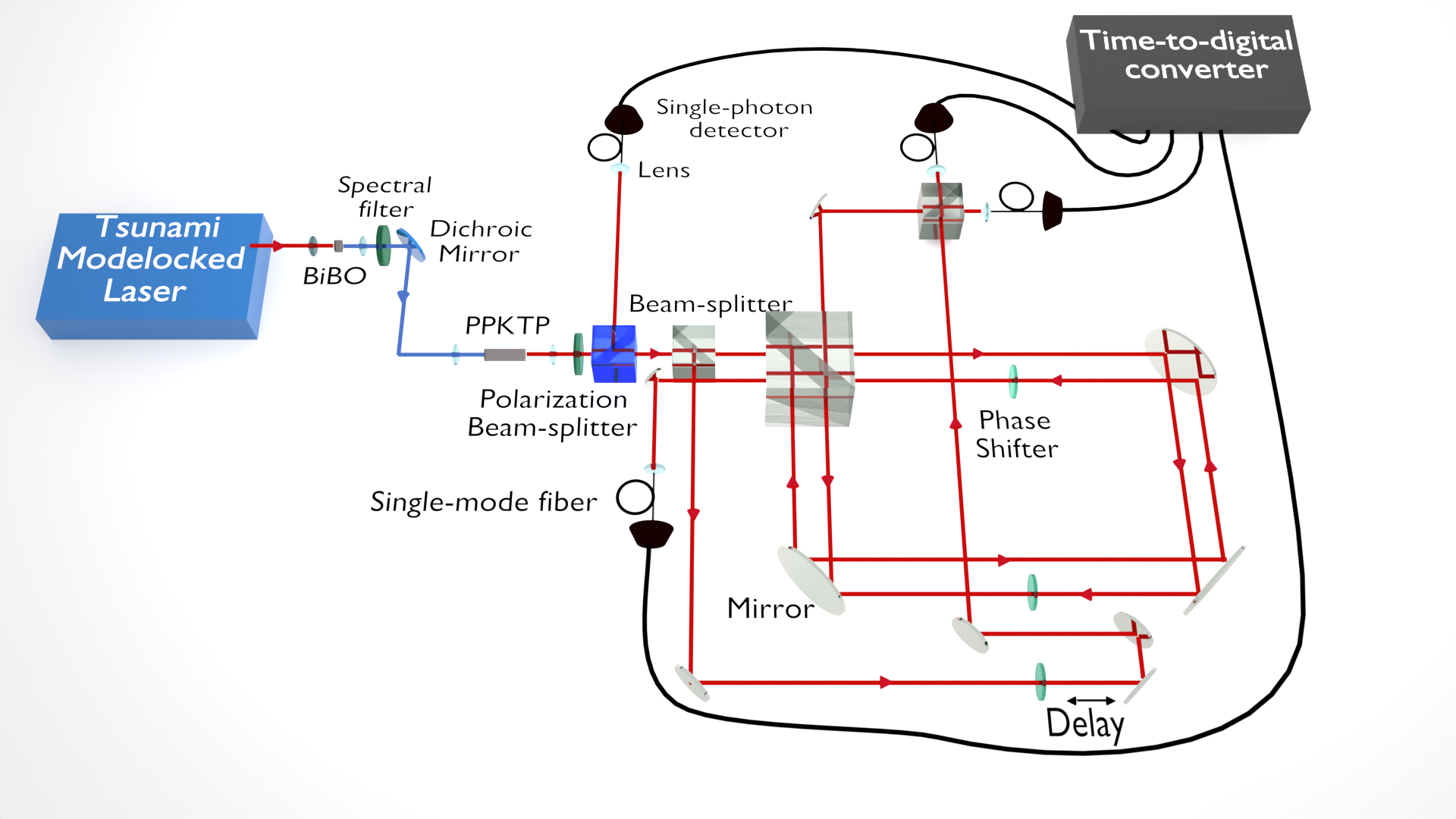}
    \caption{Experimental setup: (a) Photon pairs are generated by pumping a PPKTP crystal with the second harmonic of a pulsed laser (generated in BiBO); (b) The heralded single photons are sent to a three-port optical interferometer consisting of an inner Sagnac loop and an outer Mach Zehnder (MZ) interferometer with information encoded by auto-controlled phase plates; (c) The heralded single photons are coupled into single-mode fibers (SMF) and detected by avalanche photodiodes (APD); different combinations of coincidence counts are processed by a time-to-digital converter (TDC).}
    \label{fig:setup}
\end{figure*}

\section{Experimental demonstration of enhanced multiple-access communication using a single photon}

\label{Sect:experiment}

We have applied our communication framework to a multi-port optical interferometer experiment in which each sender controls one path that the particle can traverse.  A single photon is used as the message carrier, prepared into the desired superposition mode via the interferometric structure. Messages are coherently encoded by different senders onto the photon along each optical path of the interferometer and decoded by the single receiver, who collects the photon at the output ports of the channel.
\par Not all the communication protocols described above can be faithfully implemented using such a setup, due to various unavoidable experimental imperfections, including finite transmission and detection efficiencies, a non-ideal probabilistic single-photon source with multi-photon pair generation, and imperfect interference visibility of the optical interferometer. In particular, the photon loss incurred from the finite detection/transmission efficiency prevents us from exploring the vacuum mode as a valid decoding outcome.  Furthermore, the quantum enhancement is extremely sensitive to interferometric visibility, as we will explain in detail later. Taking all these factors into consideration, the most viable experiment to conduct is the two-sender coherence-assisted communication protocol (Proposition \ref{prop:two-sender-assisted-protocol}) presented in Section ~\ref{Sect:general_N_encoding}. 
 The advantage of this scenario is that quantum-enhanced communication can be achieved using only phase encoding by each sender. However, as argued in Section ~\ref{Sect:general_N_encoding}, the communication rates are the same as in a two-sender unassisted protocol using path blocking and phase encoding on the uniform superposition state $\frac{1}{\sqrt{2}}(\ket{\mbf{e}_1}+\ket{\mbf{e}_2})$ (see the solid line in Fig. \ref{Fig:2-sender-full-quantum-region}).
 
 The experimental setup for this protocol is shown in Fig.~\ref{fig:setup}. A heralded single photon is created from spontaneous parametric down conversion (SPDC), and sent to a three-port interferometer with splitting ratio $1/2:1/4:1/4$. The single-photon state is filtered with a polarizer and spectral filter and coupled into single-mode fiber (SMF), which allows us to ignore all of its internal degrees of freedom and write down the corresponding heralded state as a superposition of different path basis states $\ket{\mbf{e}_i} = \ket{0}^\msf{A_1}\cdots\ket{1}^{\msf{A}_i}\cdots\ket{0}^{\msf{A}_N}$ :
\begin{equation}
    \ket{\psi}=\frac{1}{\sqrt{2}}\ket{\mbf{e}_1}+\frac{1}{2}\ket{\mbf{e}_2} +\frac{1}{2}\ket{\mbf{e}_3},
\end{equation} 
where the third path is the assistance path, while senders 1 and 2 each encode their input bits onto the photon locally with tunable phase shifters in the form of glass windows. The phase shifters are characterized with respect to the angle of rotation of the glass window and a phase-shift of 0 is set to encode the bit ``0'' and $\pi$ to encode
the bit ``1''. At the output ports of the interferometric setup, single-photon detectors are placed and information is decoded purely based on the which-port information.

\subsection{Experimental results}
In order to claim the implementation of a communication protocol with only one single particle involved, we characterize the heralded second-order cross-correlation function at zero delay $g_{hcc}^{(2)}(0)$ of our photon-pair source. For an ideal source this number should be 0, which means exactly one photon is produced in a heralded manner; however, without a perfect photon-number-resolving detector, there will always be a trade-off between having a higher heralded-single-photon rate and lower $g_{hcc}^{(2)}(0)$. We measure $g_{hcc}^{(2)}(0)$ = $0.0017 \pm 0.001$, which can basically rule out the possibility of having more than one particle traveling into the communication setup after heralding. This value is set to be an order of magnitude smaller than our expected quantum violation, as we will elaborate on later.
 \par
\begin{figure}[h]
     %\centering
    \includegraphics[width=0.5\textwidth]{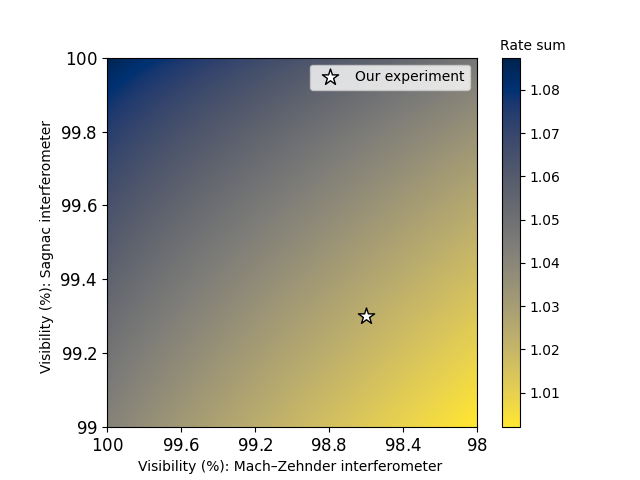}
    \caption{Expected enhanced two-sender communication rate as a function of the interference visibilities of the inner Sagnac and outer Mach-Zender (MZ) interferometers. The maximal capacity rate of $1.0875$ is achieved when perfect interference visibility is obtained.}
   \label{visibility}
\end{figure}

 
A non-ideal single-photon source with small multi-photon, encoding operations, and/or decoding detections can all degrade the performance of our quantum protocol to some extent. Among them, most error in the setup is actually caused by the non-unit interference visibility. Ideally, when the three-port optical interferometer has perfect interference visibility the following transition probabilities can be achieved:
\begin{align}
p(1|10)&=1, &p(2|11)&=1,\notag\\
p(0|00)&=\frac{1}{2},&p(1|00)&=\frac{1}{4},&p(2|00)&=\frac{1}{4}\notag\\
p(0|01)&=\frac{1}{2},&p(1|01)&=\frac{1}{4},&p(2|01)&=\frac{1}{4}.\label{Eq:transition-balanced}
\end{align}
%, a transition probability of the form of Eq.~ (\ref{eq:two-sender-unassisted-equal-superposition-channel}) can be recovered and a maximal mutual information of $I(X_1X_2:Y)=1.0875$ can be obtained. 
However, the communication protocol is extremely sensitive to the interference visibility, as shown in Fig.~\ref{visibility}. To obtain a greater quantum enhancement with better interferometric visibility, we devise a three-port optical interferometer comprised of a passively stabilized Mach-Zehnder (MZ) interferometer with an offset Sagnac interferometer embedded within it. The visibility of the Sagnac interferometer is achieved to above $V_s=99.5\pm 0.2\%$ after tightly filtering the single photon spectrally and spatially, and the visibility for the outer MZ interferometer is around $V_z=98.2\pm 0.4\%$ averaged over 10 minutes.%, which together give an estimated quantum enhancement $I(X_1X_2:Y)_{\text{est}}\approx 1.018$. 

Our experimental demonstration of quantum advantage comes in two forms.  We first build a channel having transition probabilities close to those of Eq. \eqref{Eq:transition-balanced}.  With this channel, it is, in principle, possible to achieve asymptotic communication rates strictly larger than what is possible using a classical particle.  Second, we go one step further and actually use the channel to establish correlated random variables between the senders and receivers whose mutual information is above one, thereby exceeding the accessible information of a classical particle.

\begin{figure*}[t]
    \centering
    \includegraphics[width=\textwidth]{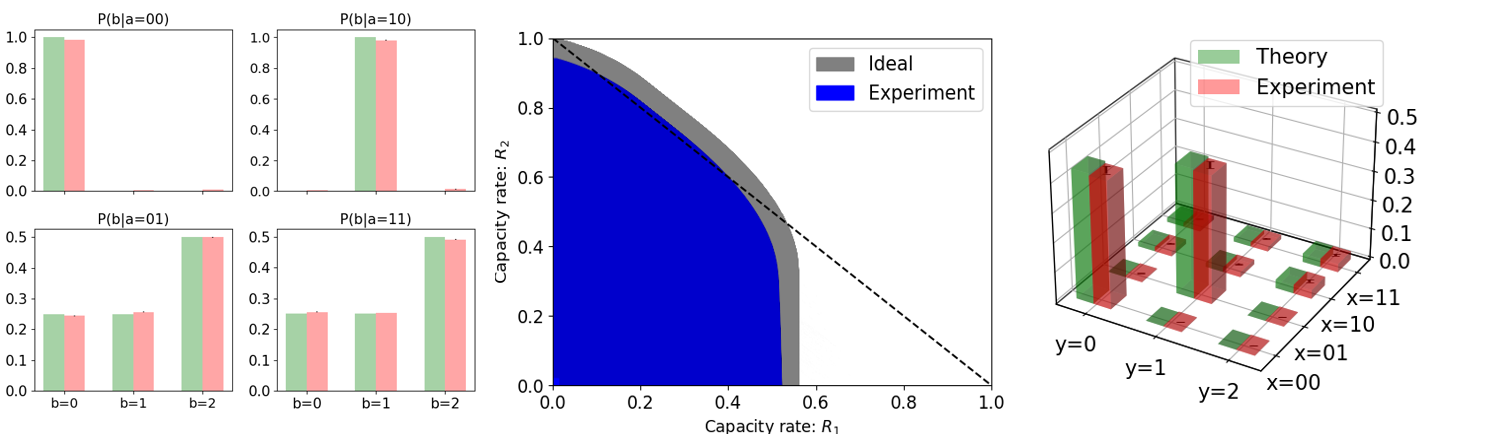} 
    \caption{(a) Example of transition probability  $p(y|x)$ from direct characterization of the two-sender channel where inputs $\bm{x}=(x_1,x_2)\in\{0,1\}^{\times 2}$ and output $y\in\{0,1,2\}$ (b) The union of achievable rate regions with the corresponding channel, with experiment in blue and the ideal case in grey. The dashed line represents the bound of the rate region achievable by a classical MAC. (c) Comparing our empirical joint distribution $p(\bm{x},y)_{\text{empirical}}$ to its theoretical value for two-sender channel inputs $\bm{x}\in\{0,1\}^{\times 2}$ and output $y\in\{0,1,2\}$ where error bars are statistical uncertainty.}
    \label{fig:transition probability}
\end{figure*}
\subsubsection{Characterizing a two-sender assisted channel by transition probabilities}
To demonstrate quantum enhancement in the two-sender communication protocol, we first characterize the transition probability of the channel $p(y|\bm{x}=(x_1,x_2))$, where $x_{i}$ is the bit encoded by sender $i$ corresponding to $0$ ($\pi$) phase for $x_{i}=0$ ($x_{i}=1$), while $y$ is the trit decoded by the receiver based on the ``which-port'' information of the output particle measured.  Given the low $g_{hcc}^{(0)}$ we set, we characterize each transition probability with different inputs $\bm{x}=(x_1,x_2)$ by registering coincident counts over a three minute period with around $N=10^5$ events registered (see Fig.~\ref{fig:transition probability} (a)).
\par 
Using the measured transition probabilities $p(y|\bm{x})$, the asymptotic rate region for the constructed channel can be computed.  In particular, for the ideal channel of Eq. \eqref{Eq:transition-balanced}, the mutual information between senders and receivers is found to be maximized by a uniform prior distribution for $x_1\in\{0,1\}$ and a biased distribution for $x_2\in\{0,1\}$ with $\text{Pr}\{x_2=0\}\approx \frac{15}{17}$.  With $X=(X_1,X_2)$ denoting input variables with these distributions, our constructed channel can thus achieve an input-output mutual information of  
   \begin{align}
   I(X:Y)_{\text{ch}} &= \sum_{x,y} p(x)p(y|x)\log\frac{p(y|x)}{p(x)}\notag\\
   &=1.0152\pm 0.0034,
\end{align} 
where the error is the standard deviation over 10 runs of the experiment to take both statistical and systematic error into consideration (the estimation on the statistical error is given in the supplementary material).  More generally, by varying the prior $p(\bm{x})=p(x_1)p(x_2)$, a different rate region is determined by the three mutual information quantities $\{I(X_1:Y|X_2), \;I(X_2:Y|X_1), \;I(X_1X_2:Y)\}$ via Eq.~(\ref{eq:rateregion_constrian}). The union of these regions is presented in Fig.~$\ref{fig:transition probability}$ (b). 
\begin{comment}
\begin{figure}[h]
    \centering
    \includegraphics[width=0.5\textwidth]{plt_tran.png}
    \caption{, where error includes statistical error in photon counts as well as error in Random bits generation.}
    \label{fig:joint probability}
\end{figure}
\end{comment}
\subsubsection{Characterizing a two-sender assisted channel by mutual information}
\label{Sect:experiment-mutual-information}
We take the demonstration further by generating empirical random variables $(X,Y)$ that are correlated using the single-particle channel we build.  Ideally we would like their mutual information $I(X:Y)$ to be close to the maximum accessible information $R(\mc{Q}_2^{\text{ass}})$, but any value larger than one will already yield a quantum advantage.  To this end, we generate multiple series of random bits each of length $680$ by independently sampling from the input set $\{0,1\}$ with uniform probability $p(0)=1/2$ for input $x_1$ and biased probability $p(0)=\frac{15}{17}$ for input $x_2$.  Ideally, each sample would correspond to a specific choice of encoding in one run of the experiment.  However, in practice we can only change the encoding map once per second.  Hence, the ensemble we generate has the form $\{p(\bm{x}),\sigma_{\bm{x}}^{\otimes m}\}^{\otimes n}$ rather than (ideally) $\{p(\bm{x}),\sigma_{\bm{x}}\}^{\otimes mn}$, where both $n=680$ and $m\approx 600$ to be the coincident count rates.  Even if we assume that the decoder does not try to exploit this block structure (see the discussion on loopholes below), there are still two sources of uncertainty in this setup: (a) the generation of the random bit and (b) the photon number fluctuation in each run of the experiment.  The result is a mutual information with larger uncertainty and larger bias than $I(X:Y)_{\text{ch}}$, yet still above the classical threshold:
   \begin{align}
    I(X:Y)_{\text{empirical}} &= \sum_{x,y} p(x,y)\log\frac{p(x,y)}{p(x)p(y)}\notag\\
    &=1.0117\pm 0.0047,
\end{align} 
where again the error is the standard deviation over 10 runs of the experiment. Here, $I(X:Y)_{\text{empirical}}$ is computed using the empirical joint distribution $p(x,y)_{\text{empirical}}$ compiled from both the input and output data.  %Given the prior distribution $p(x)=p(x_1)p(x_2)$ of the random input bits, each the rate region is determined by the three mutual information quantities $\{I(X_1:Y|X_2), \;I(X_2:Y|X_1), \;I(X_1X_2:Y)\}$ via Eq.~(\ref{eq:rateregion_constrian}). The achievable rate regions are shown in Fig.~$\ref{fig:joint probability}$(b). 

\subsection{Experimental imperfections and loopholes}
Similar to the problems encountered in most photonic Bell tests \cite{Christensen2013,Brunner-2014a,Collaboration2018}, our communication framework suffers from several experimental loopholes.  While these can be fixed in principle, they make an experimental demonstration of enhanced quantum communication challenging to attain at the single-particle level.  

\subsubsection{Detection loophole}
\begin{figure}
    \centering
    \includegraphics[width=0.5\textwidth]{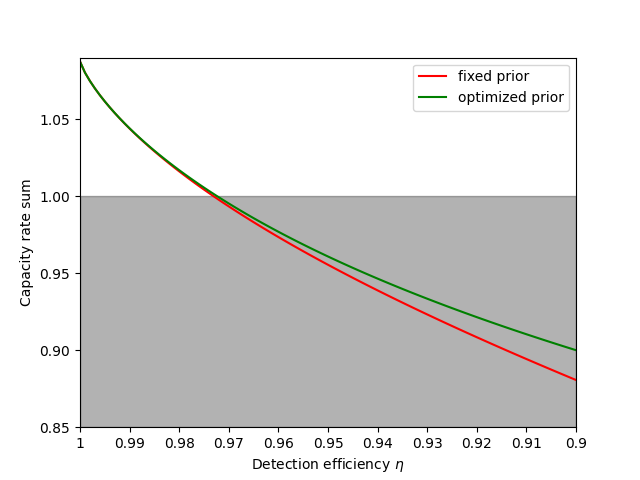}
    \caption{For a non-ideal single photon detector, no quantum enhancement can be observed in the two-sender unassisted scenario when the detector efficiency $\eta$ drops below roughly $97\%$ (and all other apparatuses behave flawlessly).}
    \label{fig:detectionloophole}
\end{figure}

In optical experiments, the main difficulty in demonstrating our theoretical protocols is the limited photon detection efficiency, which generates many ``no-click'' events.  The single-photon detector we employ (APD, Excelitas SPCM-AQ4C) has a photon detection efficiency around $40\%$ at our working wavelength of $810$ nm. This ratio can be improved up to $95\%$ with superconducting single-photon detectors.  Yet, even this relatively high efficiency is insufficient to implement many single-particle communication protocols.  The standard way of demonstrating a detection-loophole free Bell test is to classically relabel no-click events as some other detection event.  Unfortunately, this is not a good strategy in any communication protocol that uses blocking as an encoding operation since then ``no-click'' events are intentionally used to transmit information.  To see this quantitatively, consider the two-sender unassisted protocol in Section~\ref{sect:two-sender-MAC} that uses blocking as an encoding operation.  When starting with a uniform superposition $\frac{1}{\sqrt{2}}(\ket{\mbf{e}_1}+\ket{\mbf{e}_2})$ and following the encodings of Eq. \eqref{eq:two-sender-2,2-input-encoding}, the resulting channel without detection efficiency has the transition probabilities of Eq. \eqref{Eq:transition-balanced}.  If we assume the detection efficiency is a constant $\eta$ for all detectors, then the transition probabilities are replaced by
\begin{align}
    &p(1|10)=\eta,\quad p(0|10)=1-\eta\notag,\\
    &p(2|11)=\eta,\quad p(0|11)=1-\eta\notag,\\
    &p(0|00)=1-\frac{1}{2}\eta,\quad p(1|00)=\frac{1}{4}\eta,\quad p(2|00)=\frac{1}{4}\eta\notag,\\
    &p(0|01)=1-\frac{1}{2}\eta,\quad p(1|01)=\frac{1}{4}\eta,\quad p(2|01)=\frac{1}{4}\eta.\notag
\end{align}
As shown in Fig.~\ref{fig:detectionloophole}, the largest capacity rate sum of this channel drops below one quickly.
A similar situation occurs if the transmission efficiency is low (below $97\%$ in the above case), which is almost inevitable in optical experiments. 
\par 
This experimental imperfection leads to two consequences.  First, we cannot perform any protocol with block operations using our current technologies. Second, even for the case of using just phase encoding, our experiment does not close the detection loophole but instead uses the assumption of ``fair sampling.''  In other words, we assume that the accepted data in our experiment is representative of the data that would have been recorded if the detectors had unit efficiency \cite{Christensen2013}.

\subsubsection{Freedom-of-choice loophole}
The freedom-of-choice loophole has recently been proposed and fixed in Bell tests \cite{Brunner-2014a, Collaboration2018}.  This loophole refers to the possibility that ``hidden variables'' may influence the choice of measurements in experiments and thus enable cheating in acquiring the empirical results. 
\par
A similar concern could also be raised in the experimental demonstration presented in Section \ref{Sect:experiment-mutual-information}.  As described, the time delay in our ability to switch the encoding of each sender means that the same channel input is selected in $m=O(10^3)$ consecutive experimental runs.  This lack of input freedom for each trial could be exploited in some classical protocol that is attempting to reproduce the same mutual information $I(X:Y)_{\text{empirical}}>1$.
%Consider the case where the senders choose the same input repetitively: based on the result for the first few particles, the receiver can adjust their channel to make sure they receive a more 'distinguishable' outcome, thereby achieve a higher capacity at the end. In our first experiment, participants can cheat when characterizing the channel with those fixed inputs. By building different channels given different inputs, a 'channel' that can send more information can be faked. 
The ultimate way of fixing this problem is to independently choose an input $(x_1,x_2)$ and apply encoding $\mc{E}^{\msf{A}_1}_{x_1}\otimes\mc{E}^{\msf{A}_2}_{x_2}$ %and measure the bivariate distribution $p(x,y)$, which we partially implement in our second experiment but not in a complete manner. In principle, the random encoding should be chosen per 
for each incoming photon.  This requires a phase encoding operation as fast as $80$ MHz in order to match our laser repetition rate. This can be achieved potentially with electro-optic devices or acousto-optical devices; however, due to the demanding requirements of the overall interference visibility for the interferometric setup, we could not easily introduce such components into our setup.  %thus, this loophole is still open.  

\subsubsection{Locality loophole.}
In standard Bell experiments, the locality constraint is set to prevent the two sites from communicating with each other \cite{Massa2019}.  %, where the experimenters can cheat on the result if their measurement strategies can be correlated, while in the two-way communication scenario, the situation is slightly different where they have to  where they ensure that photon cannot be exchanged more than once between the two parties \cite{Massa2019}. \par 
Our experiment has a similar loophole in that without sufficient separation between the senders, it is potentially possible for them to communicate and perform some joint (i.e. not independent) encoding on the particle.  %suffers similar locality loopholes, in which we have to prevent the photon from passing two senders in a consecutive way, otherwise, information can be encoded consecutively by different senders. 
To avoid this, at least we would need to design the experiment so that the communication time between senders is much longer than the time it takes the photon to travel from one sender to the receiver. In our case, the required time difference is determined by the coincidence window we set, which requires the spatial separation between senders to be greater than $2\text{ns}\times c=0.6\text{m}$. Closing this loophole in our setup is challenging since the overall interference visibility and stability are limited by the size of the interferometer.

\section{Conclusion}

In the present paper, we investigated how much information can be transmitted from multiple senders to a single receiver by the use of a single quantum or classical particle. To analyze this question and to show an advantage of quantum over classical particles, we have created a framework of classical multiple-access channels constructed by locally modulating an initial superposition state of different paths and afterwards detecting the particle with a general measurement. The classical case is included 
 when the initial state assigns a definite path to the particle; on the other hand, if the initial state is a genuine quantum superposition, 
it has the potential to induce channels not reachable with a classical state. 
 Specifically, we found that the communication rates of independent messages of the separate users show a clear quantum advantage.  Indeed,  
 for a single classical particle, the rate-sum for any number of senders is bounded by one, while it exceeds one for two or more sender, being 
 monotonically increasing in the number of senders. The rate-sum can be even larger in the model of coherence assistance, where there is 
 another path from the source directly to the decoder, which allows a rate exceeding 1 even for the single-sender model, to be precise 
 1.0931 bits per channel use. We also experimentally demonstrated our 
predicted quantum advantage by implementing the two-sender coherence-assisted 
protocol using an optical interferometric setup. The constructed channel supports a communication rate-sum of $1.0152\pm0.0034$, showing a four-standard-deviation quantum advantage over the classical bound.  Alternatively, the channel can be used to correlate random variables whose mutual information we empirically found to be $I(X:Y)_{\text{empirical}}=1.0117\pm 0.0047$, again exceeding the classical threshold of one.

 We leave a number of open questions regarding the basic theoretical understanding of the single-particle MAC, starting with the actual maximum value of the rate-sum for any number of senders and the characterization of the full capacity region. Our best upper 
 bound on the rate-sum is the Holevo quantity, and while we are just 
 short of calculating that exactly, it scales as $\log N$ for large 
 number $N$ of senders. By contrast, we do not even know if the achievable 
 rate-sum via accessible information diverges or not. It seems we would want better outer bounds on the capacity region, but it is 
 perhaps much more exciting to search for improved modulation and detection schemes. In another direction, fixing the particular initial state, but optimizing over modulations and detection, the  achievable rate region could give new quantifiers for the amount of coherence in the state along the lines of Refs. \cite{Biswas-2017a, Napoli-2016a, Baumgratz-2014a}.

\section*{Acknowledgements}
This work was supported by the National Science
Foundation Award Nos. 1839177 and 2112890.  AW is supported by the European Commission QuantERA grant ExTRaQT (Spanish MICINN project PCI2022-132965), by the Spanish MINECO (project PID2019-107609GB-I00) with the support of FEDER funds, the Generalitat de Catalunya (project 2017-SGR-1127), by the Spanish MCIN with funding from European Union NextGenerationEU (PRTR-C17.I1) and the Generalitat de Catalunya, and by the Alexander von Humboldt Foundation, as well as the Institute of Advanced Study of the Technical University Munich.

\bibliographystyle{unsrt}
\bibliography{MACs}

\appendix
\onecolumngrid
\newpage

\section{Lower bound for one-sender ($R(\mc{Q}_1^*)$)}
\label{sect:one-sender-appendix}
Using the encoding operations given in Eq.~\eqref{Eq:one-sender-encoding}, the encoded cq state is
\begin{align*}
    \sigma^{\msf{XAR}} &= (1-q)\state{0}\otimes\sigma_0 + \frac{q}{2}\state{1}\otimes\sigma_1 + \frac{q}{2}\state{2}\otimes\sigma_2 \notag\\
    &= (1-q)\state{0}\otimes \left(\cos^2\theta\ket{00}\bra{00}+\sin^2\theta\ket{\mbf{e}_2}\bra{\mbf{e}_2}\right) \notag\\
    &\quad+ \frac{q}{2}\state{1}\otimes \left(\cos\theta\ket{\mbf{e}_1}+\sin\theta\ket{\mbf{e}_2}\right) \left(\cos\theta\bra{\mbf{e}_1}+\sin\theta\bra{\mbf{e}_2}\right) \notag\\
    &\quad+ \frac{q}{2}\state{2}\otimes \left(e^{i\alpha}\cos\theta\ket{\mbf{e}_1}+\sin\theta\ket{\mbf{e}_2}\right) \left(e^{i\alpha}\cos\theta\bra{\mbf{e}_1}+\sin\theta\bra{\mbf{e}_2}\right).
\end{align*}
To calculate its accessible information, we first note that the optimal POVM achieving the accessible information can be taken to be rank-1 projectors \cite{Davies-1978}. Additionally, as noted in the main text, the ensemble has the following symmetries: (i) $\sigma^{\msf{XAR}}$ is diagonal in the number basis, and  (ii) $q/2\cdot \sigma_1$ and $q/2\cdot \sigma_2$ are related by a reflection across the line $y=x\tan(\alpha/2)$ in the $x-y$ plane of the Bloch sphere.  Using the same arguments in Ref. \cite{Frey-2006} (Proposition 1), we deduce that the optimal measurement attaining the accessible information can be made to have the same symmetries. Therefore, the optimal POVM can be taken to be $\{\state{00},w_m\state{\pi_m},\,w_m\state{\pi_m'}\}$,
where 
\begin{align}
    &\ket{\pi_m}=\sqrt{\overline{\sigma_m}}\ket{\mbf{e}_1}+\sqrt{\sigma_m}e^{i\beta_m}\ket{\mbf{e}_2}   \\
    &\ket{\pi_m'}=\sqrt{\overline{\sigma_m}}\ket{\mbf{e}_1}+\sqrt{\sigma_m}e^{-i(\alpha+\beta_m)}\ket{\mbf{e}_2}.
\end{align}
Here $\ol{\sigma_m}=1-\sigma_m$. Each $m$ labels a pair of symmetric projectors specified by $(w_m,\sigma_m,\beta_m)$. Now, since $\sum_m (\Pi_m+\Pi_m')$ is the projector onto the $\ket{\mbf{e}_1},\ket{\mbf{e}_2}$ subspace, we have 
\begin{align}
    \sum_m \left[w_m \begin{pmatrix}
    \ol{\sigma_m} &\sqrt{\ol{\sigma_m}\sigma_m}e^{-i\beta_m} \\
    \sqrt{\ol{\sigma_m}\sigma_m}e^{i\beta_m} &\sigma_m \\
    \end{pmatrix} + 
    w_m \begin{pmatrix}
    \ol{\sigma_m} &\sqrt{\ol{\sigma_m}\sigma_m}e^{i(\alpha+\beta_m)} \\
    \sqrt{\ol{\sigma_m}\sigma_m}e^{-i(\alpha+\beta_m)} &\sigma_m \\
    \end{pmatrix}\right] = \mbb{I},
\end{align}
from which we can conclude that 
\begin{align}
    \sum_m w_m\sigma_m = \frac{1}{2}, \qquad \sum_m w_m =1, \qquad \sum_m w_m\sqrt{\ol{\sigma_m}\sigma_m}(e^{i\beta_m}+e^{-i(\alpha+\beta_m)})=0.\label{Eq:sigma-beta-constraint}
\end{align}
Denote the set of $\{(w_m,\sigma_m,\beta_m)\}_m$ satisfying all three constraints in Eq.~\eqref{Eq:sigma-beta-constraint} as $\mc{S}$. Following the same approach laid out in \cite{Frey-2006}, the accessible information of the ensemble (and hence the total communication rate) is given by 

\begin{align}
    I_{acc}=\max_\mc{S}\sum_m w_m J(\sigma_m,\beta_m;q,\theta,\alpha),
\end{align}
where 
\begin{align}
    J(\sigma,\beta;q,\theta,\alpha)=&q|\sqrt{\ol{\sigma}}\cos\theta+e^{i\beta}\sqrt{\sigma}\sin\theta|^2\log|\sqrt{\ol{\sigma}}\cos\theta+e^{i\beta}\sqrt{\sigma}\sin\theta|^2 \notag\\ 
    &+ q|\sqrt{\ol{\sigma}}\cos\theta+e^{i(\beta-\alpha)}\sqrt{\sigma}\sin\theta|^2\log|\sqrt{\ol{\sigma}}\cos\theta+e^{i(\beta-\alpha)}\sqrt{\sigma}\sin\theta|^2 \notag \\
    &+ 2(1-q)\sigma\sin^2\theta\log(\sigma\sin^2\theta) \notag \\
    &- 2\kappa\log\kappa - (1-q)\cos^2\theta\log(1-q),
\end{align}
in which $\kappa = q\ol{\sigma}\cos^2\theta+q\left[\cos\beta+\cos(\beta-\alpha)\right]\sqrt{\ol{\sigma}\sigma}\cos\theta\sin\theta+\sigma\sin^2\theta$.

We can relax the restriction on $w_m$, $\sigma_m$ and $\beta_m$ by dropping the last condition in Eq.~\eqref{Eq:sigma-beta-constraint}, thus obtaining an upper bound. Formally, let $\tilde{\mc{S}}$ denote the set of $\{(w_m,\sigma_m,\beta_m)\}_m$ that satisfy only the first two conditions in Eq.~\eqref{Eq:sigma-beta-constraint}, then
\begin{align}
    \widetilde{I_{acc}}=\max_{\tilde{\mc{S}}}\sum_m w_m J(\sigma_m,\beta_m;q,\theta,\alpha)\geq I_{acc}.
\end{align}

Note that dropping the third condition essentially allows us to optimize $\beta_m$'s freely independent of any other parameter. Our first goal is to find the optimal phase encoding $\alpha$, denoted by $\alpha^*$, that maximizes the function $J(\sigma,\beta;q,\theta,\alpha)$.

\begin{lemma}\label{lemma:optimal-alpha-beta}
For any $\sigma$, $q$, and $\theta$, $J(\sigma,\beta;q,\theta,\alpha)$ is maximized only if $(\alpha,\beta)=(0,0),\,(0,\pi),\,(\pi,0)$ or $(\pi,\pi)$.
\end{lemma}

\begin{proof}
For $J$ to attain a local maximum, it is necessary that the directional derivative $D_{\vec{u}}J=0$ and the second directional derivative $D^2_{\vec{u}}J\leq 0$ along any direction $\vec{u}$ on the $\alpha$-$\beta$ plane. Specifically, let us consider two direction given by $\vec{u_1}=(1,0)^\intercal$ and $\vec{u_2}=(1,1)^\intercal$. Then we have: 
\begin{equation}
    D_{\vec{u_1}}J=\frac{\partial J}{\partial\alpha}=0,\quad
    D_{\vec{u_2}}J=\frac{\partial J}{\partial\alpha}+\frac{\partial J}{\partial\beta}=0,
\end{equation}
\begin{equation}
    D^2_{\vec{u_1}}J=\frac{\partial^2 J}{\partial\alpha^2}\leq 0,\quad
    D^2_{\vec{u_2}}J=\left(\frac{\partial}{\partial\alpha}+\frac{\partial}{\partial\beta}\right)\left(\frac{\partial J}{\partial\alpha}+\frac{\partial J}{\partial\beta}\right) \leq 0.
\end{equation}
Calculating the first derivatives gives:
\begin{align}
    D_{\vec{u_1}}J
    &= \frac{1}{\ln{2}}q\left(\ln|\sqrt{\ol{\sigma}}\cos\theta+e^{i(\beta-\alpha)}\sqrt{\sigma}\sin\theta|^2+1\right)2\sqrt{\ol{\sigma}\sigma}\cos\theta\sin\theta\sin(\beta-\alpha) \notag\\
    &\hspace{24pt}-\frac{1}{\ln{2}}2(\ln\kappa+1)q\sqrt{\ol{\sigma}\sigma}\cos\theta\sin\theta\sin(\beta-\alpha) = 0, \\
    D_{\vec{u_2}}J
    &= -\frac{1}{\ln{2}}q\left(\ln|\sqrt{\ol{\sigma}}\cos\theta+e^{i\beta}\sqrt{\sigma}\sin\theta|^2+1\right)2\sqrt{\ol{\sigma}\sigma}\cos\theta\sin\theta\sin\beta \notag\\
    &\hspace{24pt}+\frac{1}{\ln{2}}2(\ln\kappa+1)q\sqrt{\ol{\sigma}\sigma}\cos\theta\sin\theta\sin\beta = 0.
\end{align}
Assuming $q\sqrt{\ol{\sigma}\sigma}\cos\theta\sin\theta\ne0$ (when one of $q$, $\cos\theta$, and  $\sin\theta$ is 0, the ensemble becomes trivial, and when one of $\sigma$ and $\ol{\sigma}$ is 0, then $J$ reduces to $-(1-q)\cos^2\theta\log(1-q)$, which is independent of $\alpha$ and $\beta$), the two equations simplify to:
\begin{equation}
    \begin{cases}
    \sin(\beta-\alpha)\log|\sqrt{\ol{\sigma}}\cos\theta+e^{i(\beta-\alpha)}\sqrt{\sigma}\sin\theta|^2 -\sin(\beta-\alpha)\log\kappa=0 \\
    \sin\beta\log|\sqrt{\ol{\sigma}}\cos\theta+e^{i\beta}\sqrt{\sigma}\sin\theta|^2 -\sin\beta\log\kappa=0 \\
    \end{cases}
\end{equation}
This set of equations admits four possible conditions:
\begin{align}
    &\text{(i) }\sin(\beta-\alpha)=0,\;\sin\beta=0;\\
    &\text{(ii) }|\sqrt{\ol{\sigma}}\cos\theta+e^{i(\beta-\alpha)}\sqrt{\sigma}\sin\theta|^2=\kappa,\;\sin\beta=0;\\
    &\text{(iii) }\sin(\beta-\alpha)=0,\;|\sqrt{\ol{\sigma}}\cos\theta+e^{i\beta}\sqrt{\sigma}\sin\theta|^2=\kappa;\\
    &\text{(iv) }|\sqrt{\ol{\sigma}}\cos\theta+e^{i(\beta-\alpha)}\sqrt{\sigma}\sin\theta|^2=\kappa,\;|\sqrt{\ol{\sigma}}\cos\theta+e^{i\beta}\sqrt{\sigma}\sin\theta|^2=\kappa.
\end{align}
\begin{comment}
with solutions:
\begin{align}
    &\text{(i) } (\alpha,\beta)=(0,0),\,(0,\pi),\,(\pi,0),\,\text{or}\,(\pi,\pi) \\
    &\text{(ii) } (\alpha,\beta)=\left(\cos^{-1}\left[\frac{1-q}{q}\sqrt{\frac{\ol{\sigma}}{\sigma}}\cot\theta + \frac{2-q}{q}\right],0\right)\,\text{or}\,\left(\cos^{-1}\left[-\frac{1-q}{q}\sqrt{\frac{\ol{\sigma}}{\sigma}}\cot\theta +\frac{2-q}{q}\right],\pi\right) \\
    &\text{(iii) } (\alpha,\beta)=\left(\cos^{-1}\left[\frac{q}{2-q}-\frac{1-q}{2-q}\sqrt{\frac{\ol{\sigma}}{\sigma}}\cot\theta\right],\cos^{-1}\left[\frac{q}{2-q}-\frac{1-q}{2-q}\sqrt{\frac{\ol{\sigma}}{\sigma}}\cot\theta\right]\right)\\
    &\hspace{24pt}\,\text{or}\, \left(\cos^{-1}\left[-\frac{q}{2-q}-\frac{1-q}{2-q}\sqrt{\frac{\ol{\sigma}}{\sigma}}\cot\theta\right]+\pi,\cos^{-1}\left[-\frac{q}{2-q}-\frac{1-q}{2-q}\sqrt{\frac{\ol{\sigma}}{\sigma}}\cot\theta\right]\right)\\
    &\text{(iv) } (\alpha,\beta)=\left(0,\cos^{-1}\left[-\frac{1}{2}\sqrt{\frac{\ol{\sigma}}{\sigma}}\cot\theta\right]\right)\,\text{or}\,\left(2\cos^{-1}\left[-\frac{1}{2}\sqrt{\frac{\ol{\sigma}}{\sigma}}\cot\theta\right],\cos^{-1}\left[-\frac{1}{2}\sqrt{\frac{\ol{\sigma}}{\sigma}}\cot\theta\right]\right)
\end{align}
(assuming these values exist).
\end{comment}
Now, calculating the second derivatives gives us:
\begin{align}
    D^2_{\vec{u_1}}J &= 2q\sqrt{\ol{\sigma}\sigma}\cos\theta\sin\theta\left[-\cos(\beta-\alpha)\log\left(\frac{|\sqrt{\ol{\sigma}}\cos\theta+e^{i(\beta-\alpha)}\sqrt{\sigma}\sin\theta|^2}{\kappa}\right)\right. \notag\\
    &\hspace{12pt}\left. +\frac{1}{\ln2}\sin(\beta-\alpha)\left(\frac{2}{|\sqrt{\ol{\sigma}}\cos\theta+e^{i(\beta-\alpha)}\sqrt{\sigma}\sin\theta|^2} -\frac{q}{\kappa}\right)\sqrt{\ol{\sigma}\sigma}\cos\theta\sin\theta\sin(\beta-\alpha)\right] \\
    D^2_{\vec{u_2}}J &=2q\sqrt{\ol{\sigma}\sigma}\cos\theta\sin\theta\left[-\cos\beta\log\left(\frac{|\sqrt{\ol{\sigma}}\cos\theta+e^{i\beta}\sqrt{\sigma}\sin\theta|^2}{\kappa}\right)\right. \notag\\
    &\hspace{12pt}\left. +\frac{1}{\ln2}\sin\beta\left(\frac{2}{|\sqrt{\ol{\sigma}}\cos\theta+e^{i\beta}\sqrt{\sigma}\sin\theta|^2} +\frac{q}{\kappa}\right)\sqrt{\ol{\sigma}\sigma}\cos\theta\sin\theta\sin\beta\right]
\end{align}
Recall that $\cos\theta\sin\theta>0$ since $\theta$ can be taken to be in $[0,\pi/2]$, and we assumed $\cos\theta\sin\theta\neq0$. Plugging each of the four conditions into the two expressions above we find that $D^2_{\vec{u_1}}J>0$ for conditions (ii) and (iv), while $D^2_{\vec{u_2}}J>0$ for conditions (iii) and (iv), unless $\sin(\beta-\alpha)=\sin\beta=0$ also holds. Therefore, points satisfying (i), namely $(\alpha,\beta)=(0,0),\,(0,\pi),\,(\pi,0),\,\text{or}\,(\pi,\pi)$, are the only possible local maxima of $J$.
\end{proof}

Since we have dropped some constraints on $\beta$ and treated it as an independent variable when optimizing, the optimal $(\alpha,\beta)$ may not actually be feasible. However, it is easy to check that $(\alpha,\beta) = (0,\pi)$, $(\pi,0)$, and $(\pi,\pi)$ satisfies all of the constraints in Eq.~\eqref{Eq:sigma-beta-constraint}, and therefore they correspond to physical POVMs. This result tells us that the best phase encoding that the encoder can perform in our one-sender protocol is either $\alpha=0$ or $\alpha=\pi$. Additionally, note that $(\alpha,\beta)=(\pi,0)$ and $(\alpha,\beta)=(\pi,\pi)$ are images of each other under the reflection across $y=x\tan(\alpha/2)$. So, they correspond to the same pair of projectors, and we can freely choose either one. 

Note that, if the encoder chooses $\alpha=0$, the encoded cq state $\sigma^{\msf{XAR}}$ effectively reduces to
\begin{align*}
    \sigma^{\msf{XAR}} 
    &= (1-q)\state{0}\otimes \left(\cos^2\theta\ket{00}\bra{00}+\sin^2\theta\ket{\mbf{e}_2}\bra{\mbf{e}_2}\right) \notag\\
    &\quad+ q\state{1}\otimes \left(\cos\theta\ket{\mbf{e}_1}+\sin\theta\ket{\mbf{e}_2}\right) \left(\cos\theta\bra{\mbf{e}_1}+\sin\theta\bra{\mbf{e}_2}\right) \notag.
\end{align*}
The accessible information of this state is necessarily less than or equal to 1 bit, meaning that there is no quantum advantage. In other words, for maximal quantum advantage, one should use $\pi$ phase encoding. This is summarized by the following proposition.
\begin{proposition}
In the one-sender coherence-assisted scenario, if the encoding maps is given by Eq.~\eqref{Eq:one-sender-encoding}, then for any initial state $\ket{\psi}^{\msf{AR}}$ and any measurement POVM for $\msf{B}$, whenever there is a quantum advantage in the communication rate (i.e., whenever the communication rate exceeds 1 bit), $\alpha=\pi$ is always the optimal phase encoding that $\msf{A}$ can perform.
\end{proposition}

We can now prove the following theorem in the main text.
\begin{reptheorem}{thm:one-sender-acc-info}
There exists a one-sender coherence-assisted communication protocol that sends approximately 1.0931 bits of information, i.e., $R(\mc{Q}_1^*)\geq 1.0931$. The optimal $(q,\theta)$ that achieves this are approximately $(0.8701,\arccos(\sqrt{0.4715}))$, and the optimal measurement is the projective measurement $\{\ket{00},\frac{1}{\sqrt{2}}(\ket{\mbf{e}_1}\pm\ket{\mbf{e}_2})\}$.
\end{reptheorem}
\begin{proof}
Having established that $\alpha=\pi$ is the best encoding phase, in which case the best decoding phase is $0$ (or equivalently $\pi)$ we will set $\alpha=\beta=\pi$ and obtain
\begin{align}
    J(\sigma,\beta=\pi;q,\theta,\alpha=\pi) \equiv \tilde{J}(\sigma;q,\theta) = &q(\sqrt{\ol{\sigma}}\cos\theta+\sqrt{\sigma}\sin\theta)^2\log(\sqrt{\ol{\sigma}}\cos\theta+\sqrt{\sigma}\sin\theta)^2 \notag\\
    &+ q(\sqrt{\ol{\sigma}}\cos\theta-\sqrt{\sigma}\sin\theta)^2\log(\sqrt{\ol{\sigma}}\cos\theta-\sqrt{\sigma}\sin\theta)^2 \notag\\
    &+ 2(1-q)\sigma\sin^2\theta\log(\sigma\sin^2\theta) \notag\\
    &- 2(q\ol{\sigma}\cos^2\theta+\sigma\sin^2\theta)\log(q\ol{\sigma}\cos^2\theta+\sigma\sin^2\theta) \notag\\
    &-(1-q)\cos^2\theta\log(1-q)
\end{align}
and
\begin{align}
    I_{acc}(q,\theta)=\max_{\substack{\sum_m w_m\sigma_m=1/2 \\ \sum_m w_m=1}} \sum_m w_m \tilde{J}(\sigma_m;q,\theta)
\end{align}
Following the same argument presented in \cite{Frey-2006}, which we briefly recapitulate here for completeness, we first find that this maximization for the accessible information can be rewritten as a maximization with at most two terms \cite{hoeffding-1955}, that is,
\begin{align}
    I_{acc}(q,\theta) = \max_{\sigma_1\leq 1/2 \leq\sigma_2} \left(\frac{\sigma_2-1/2}{\sigma_2-\sigma_1}\tilde{J}(\sigma_1;q,\theta) + \frac{1/2-\sigma_1}{\sigma_2-\sigma_1} \tilde{J}(\sigma_2;q,\theta)\right)
\end{align}
which can be again rewritten as
\begin{align}
    I_{acc}(q,\theta) = \max_{\sigma_1\leq 1/2 \leq\sigma_2} \left[\tilde{J}(\sigma_1;q,\theta)+\frac{1/2-\sigma_1}{\sigma_2-\sigma_1}\left(\tilde{J}(\sigma_2;q,\theta)-\tilde{J}(\sigma_1;q,\theta)\right)\right].
\end{align}
The maxand can be understood as the value of the line through points $(\sigma_1,\tilde{J}(\sigma_1)$ and $(\sigma_2,\tilde{J}(\sigma_2)$ at $\frac{1}{2}$. For each $\theta$, we can find three different measurement regimes. When $q$ is sufficiently small, the optimal $(\sigma_1,\sigma_2)$ is (0,1), corresponding to optimal measurement vectors $\ket{\mbf{e}_1}$ and $\ket{\mbf{e}_2}$. As $q$ becomes larger, the optimal $(\sigma_1,\sigma_2)$ is 0 and some $\sigma^*\in[1/2,1]$, corresponding to a POVM with $\ket{\mbf{e}_1}$ and a mirror-symmetric pair of rank-one projectors. Finally, when $q$ is sufficiently close to 1, the optimal $(\sigma_1,\sigma_2)$ is (1/2,1/2), corresponding to projective measurements $\{\frac{1}{\sqrt{2}}(\ket{\mbf{e}_1}\pm\ket{\mbf{e}_2})\}$.

By straightforward calculation, we find that in region 1, the accessible information of the ensemble is
\begin{align}
    I_{acc,1}(q,\theta) = -(1-q)\cos^2\theta\log(1-q).
\end{align}
In region 3,
\begin{align}\label{eq:acc-info-reg-3}
    I_{acc,3}(q,\theta)=&q-qh_2\left(\frac{1+\sin2\theta}{2}\right)+(1-q)\sin^2\theta\log\sin^2\theta \notag\\
    &-(q\cos^2\theta+\sin^2\theta)\log(q\cos^2\theta+\sin^2\theta)-(1-q)\cos^2\theta\log(1-q)
\end{align}
where $h_2$ is the binary entropy function. In region 2, the calculation is more involved,
\begin{align}
    I_{acc,2}(q,\theta) = J(0)+\frac{1}{2}\frac{dJ}{d\sigma}(\sigma^*)
\end{align}
where $\sigma^*$ is determined from the fact that the tangent line of $J$ at $\sigma^*$ passes through $(0,J(0))$, in other words,
\begin{align}
    J(0)+\sigma^* \frac{\partial J}{\partial\sigma}(\sigma^*) = J(\sigma),
\end{align}
which after much algebra becomes
\begin{align}
    q\cos^2\theta\log\frac{q\ol{\sigma^*}\cos^2\theta+\sigma^*\sin^2\theta}{\ol{\sigma^*}\cos^2\theta-\sigma^*\sin^2\theta} = (1-q)\sigma^*\sin^2\theta\log\sigma^*
\end{align}
To plot the accessible information in the entire region of $(q,\theta)$, we note $I_{acc}(q,\theta)=\max_{i=1,2,3}\{I_{acc,i}(q,\theta)\}$. One can check by comparing the plot of $I_{acc,1}(q,\theta)$, $I_{acc,2}(q,\theta)$, and $I_{acc,3}(q,\theta)$ that the maximal accessible information occurs in region 3. To compute its value, we take the derivative of $I_{acc,3}$ with respect to $q$ and $\theta$ and set both to 0.
\begin{figure}[t]
    \centering
    \subfloat[]{\includegraphics[width=0.55\textwidth]{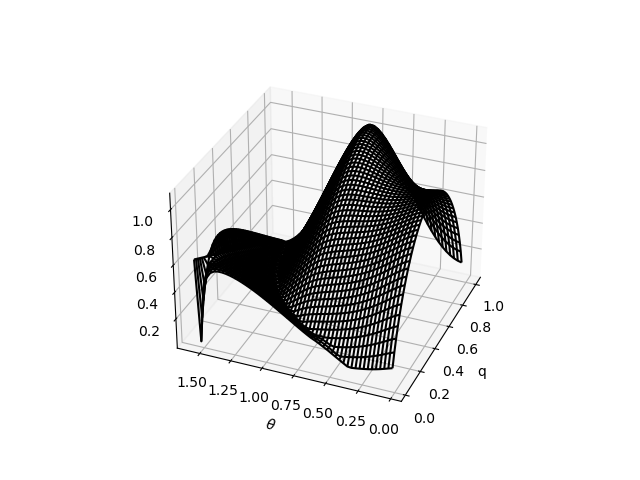}}\subfloat[]{\includegraphics[width=0.45\textwidth]{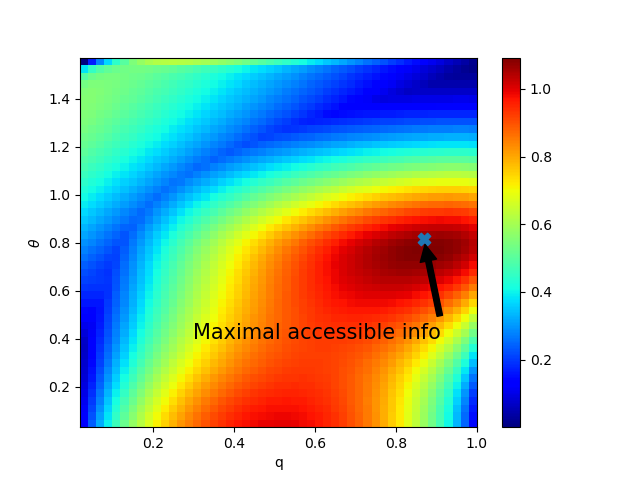}}
    \caption{The accessible information of the ensemble (assuming $\alpha=\pi$) in terms of $q$ and $\theta$.}
\end{figure}
\begin{align}
    \frac{\partial I_{acc,3}}{\partial\theta}=q\cos2\theta\log\left(\frac{1+\sin2\theta}{1-\sin2\theta}\right) + (1-q)\sin2\theta\log\left(\frac{(1-q)\sin^2\theta}{q\cos^2\theta+\sin^2\theta}\right)=0 \\
    \frac{\partial I_{acc,3}}{\partial q} = 1 - h_2\left(\frac{1+\sin2\theta}{2}\right) - \log\sin^2\theta + \cos^2\theta\log\left(\frac{(1-q)\sin^2\theta}{q\cos^2\theta+\sin^2\theta}\right)=0
    \label{Eq:dJ/dq}
\end{align}
There is no closed form solution for this system of transcendental equations. Solving these two equations numerically gives $\sin\theta^*\approx\sqrt{0.4715}$ and $q^*\approx0.8701$. This optimal choice of $\theta$ and $q$ corresponds to the intial source state $\approx\sqrt{0.4715}\ket{\mbf{e}_1}+\sqrt{0.5285}\ket{\mbf{e}_2}$, prior probability $p(x)\approx(0.1299,0.4351,0.4351)$, and the rate sum is approximately 1.0931.
\end{proof}

\begin{proposition}
If the source state is the maximally coherent state $\frac{1}{\sqrt{2}}\ket{\mbf{e}_1}+\frac{1}{\sqrt{2}}\ket{\mbf{e}_2}$, then the optimal rate is $\log(17/8)\approx1.0875>1$, and the optimal prior probabilities given by $q^* = 15/17\approx0.8824$
\end{proposition}
\begin{proof}
Take $\theta=\pi/4$ in Eq.~\eqref{eq:acc-info-reg-3} and after simplification, we find that $I(q,\pi/4) = 2q-1+h_2\left(\frac{1+q}{2}\right)$. The maximizer $q^*$ can be found by setting the derivative with respect to $q$ to 0, and we find that $q^*=15/17$, in which case the mutual information is $\log(17/8)$.
\end{proof}

\section{Lower bound for Unassisted Two-Sender ($R(\mc{Q}_2)$)}\label{sect:two-sender-appendix}
In this section, we calculate the accessible information of cq states arising from the binary-ternary encoding strategy given in Eq.~\eqref{eq:two-sender-2,3-input-encoding}. Following the same steps laid out in Sect. \ref{sect:one-sender-appendix}, the accessible information can be expressed as
\begin{align}
    I_{acc}=\max_\mc{S}\sum_m w_m J(\sigma_m,\beta_m;q,\theta,\alpha),
\end{align}
where
\begin{align}
    J(\sigma,\beta;q,q',\theta,\alpha) &=(1-q)q'(\sin^2\theta\log\sin^2\theta+2\ol{\sigma}\cos^2\theta\log\ol{\sigma}\cos^2\theta) \notag\\
    +& q(1-q')(\cos^2\theta\log\cos^2\theta+2\sigma\sin^2\theta\log\sigma\sin^2\theta) \notag\\
    +& qq'|\sqrt{\ol{\sigma}}\cos\theta+e^{i\beta}\sqrt{\sigma}\sin\theta|^2\log|\sqrt{\ol{\sigma}}\cos\theta+e^{i\beta}\sqrt{\sigma}\sin\theta|^2 \notag\\
    +& qq'|\sqrt{\ol{\sigma}}\cos\theta+e^{-i(\alpha+\beta)}\sqrt{\sigma}\sin\theta|^2\log|\sqrt{\ol{\sigma}}\cos\theta+e^{-i(\alpha+\beta)}\sqrt{\sigma}\sin\theta|^2 \notag\\
    -& \xi\log\xi - 2\eta\log\eta.
\end{align}
in which
\begin{align}
    \xi &= (1-q)(1-q')+(1-q)q'\sin^2\theta+q(1-q')\cos^2\theta = 1-q'\cos^2\theta-q\sin^2\theta\\
    \eta &= (1-q)q'\ol{\sigma}\cos\theta+q(1-q')\sigma\sin^2\theta+\frac{1}{2}qq'\left[|\sqrt{\ol{\sigma}}\cos\theta+e^{i\beta}\sqrt{\sigma}\sin\theta|^2+|\sqrt{\ol{\sigma}}\cos\theta+e^{-i(\alpha+\beta)}\sqrt{\sigma}\sin\theta|^2\right]
\end{align}
By the same argument as in Lemma \ref{lemma:optimal-alpha-beta}, we can deduce that the local extrema of function $J$ occurs only if $\alpha$ and $\beta$ are both multiples of $\pi$. And that $\alpha=\pi$ is the optimal phase encoding whenever there is a quantum advantage. Taking $\alpha=\beta=\pi$ then, we have
\begin{align}
    J(\sigma,\beta=\pi;q,q',\theta,\alpha=\pi) &=(1-q)q'(\sin^2\theta\log\sin^2\theta+2\ol{\sigma}\cos^2\theta\log\ol{\sigma}\cos^2\theta) \notag\\
    +& q(1-q')(\cos^2\theta\log\cos^2\theta+2\sigma\sin^2\theta\log\sigma\sin^2\theta) \notag\\
    +& qq'(\sqrt{\ol{\sigma}}\cos\theta-\sqrt{\sigma}\sin\theta)^2\log(\sqrt{\ol{\sigma}}\cos\theta-\sqrt{\sigma}\sin\theta)^2 \notag\\
    +& qq'(\sqrt{\ol{\sigma}}\cos\theta+\sqrt{\sigma}\sin\theta)^2\log(\sqrt{\ol{\sigma}}\cos\theta+\sqrt{\sigma}\sin\theta)^2 \notag\\
    -& \xi\log\xi - 2\eta\log\eta.
\end{align}
where
\begin{align}
    \xi &= 1-q'\cos^2\theta-q\sin^2\theta \\
    \eta &= (1-q)q'\ol{\sigma}\cos^2\theta+q(1-q')\sigma\sin^2\theta+\frac{1}{2}qq'\left[(\sqrt{\ol{\sigma}}\cos\theta-\sqrt{\sigma}\sin\theta)^2+(\sqrt{\ol{\sigma}}\cos\theta+\sqrt{\sigma}\sin\theta)^2\right] \notag\\
    &= q'\ol{\sigma}\cos^2\theta+q\sigma\sin^2\theta
\end{align}
Following the same analysis, we find three regimes for the optimal measurement, and the accessible information is maximized in the regime that corresponds to $\sigma=1/2$.
\begin{align}
    I_{acc}(q,q',\theta) = J(\sigma=\frac{1}{2},\beta=\pi;q,q',\theta,\alpha=\pi) &=(1-q)q'(\sin^2\theta\log\sin^2\theta+\cos^2\theta\log\frac{1}{2}\cos^2\theta) \notag\\
    +& q(1-q')(\cos^2\theta\log\cos^2\theta+\sin^2\theta\log\frac{1}{2}\sin^2\theta) \notag\\
    -& qq'h_2\left(\frac{1+\sin2\theta}{2}\right) - \xi\log\xi-2\eta\log\eta
\end{align}
where
\begin{align}
    \xi &= 1-q'\cos^2\theta-q\sin^2\theta \\
    \eta &= \frac{1}{2}q'\cos^2\theta+\frac{1}{2}q\sin^2\theta
\end{align}
Again taking the derivative of $I_{acc}$ with respect to $q$, $q'$, and $\theta$, we obtain the following system of equations,
\begin{align}
    \frac{\partial J}{\partial q} &= -q'(\sin^2\theta\log\sin^2\theta+\cos^2\theta\log\frac{1}{2}\cos^2\theta)
    + (1-q')(\cos^2\theta\log\cos^2\theta+\sin^2\theta\log\frac{1}{2}\sin^2\theta) \notag\\
    -& q'h_2\left(\frac{1+\sin2\theta}{2}\right) +\sin^2\theta\log\frac{\xi}{\eta} \\
    &= (2q'-1)h_2(\sin^2\theta)+q'-\sin^2\theta-q'h_2\left(\frac{1+\sin2\theta}{2}\right)+\sin^2\theta\log\frac{\xi}{\eta} = 0 \\
    \frac{\partial J}{\partial q'} &= (1-q)(\sin^2\theta\log\sin^2\theta+\cos^2\theta\log\frac{1}{2}\cos^2\theta) 
    - q(\cos^2\theta\log\cos^2\theta+\sin^2\theta\log\frac{1}{2}\sin^2\theta) \notag\\
    -& qh_2\left(\frac{1+\sin2\theta}{2}\right) +\cos^2\theta\log\frac{\xi}{\eta} \\
    &= (2q-1)h_2(\sin^2\theta)+q-\cos^2\theta-qh_2\left(\frac{1+\sin2\theta}{2}\right)+\cos^2\theta\log\frac{\xi}{\eta} = 0 \\
    \frac{\partial J}{\partial \theta} &=(q+q'-2qq')\log\tan^2\theta + qq'\cot2\theta\log\frac{1+\sin2\theta}{1-\sin2\theta} = 0
\end{align}
Numerically solving this system of equations, we obtain that the optimal $q$, $q'$, and $\theta$ is $(0.9197,0.9197,\pi/4)$, and the optimal rate sum is $1.10138$.

\section{One-sender assisted Holevo information ($\chi(\mc{Q}_1^*)$) - proof of Theorem \ref{thm:one-sender-holevo}}
\label{sect:holevo-appendix}
\begin{reptheorem}{thm:one-sender-holevo}
\begin{align}\chi(\mc{Q}_1^{\text{ass}}) &= \max_{q,\cos^2\theta\in[0,1]} q h_2(\cos^2\theta)+\cos^2\theta h_2(q) \notag\\
   &= \max_{x\in[0,1]} 2xh_2(x) \notag\\
   &\approx 1.2339,\notag
   \end{align}
\end{reptheorem}
\begin{proof}
We first show that the encoding given by Eq.~\eqref{Eq:one-sender-encoding} is in fact the best encoding strategy. Consider the most general encoding strategy using NPE operations. By convexity of the mutual information with respect to the underlying channel, it is sufficient for us to consider pure initial state $\cos\theta\ket{\mbf{e}_1}+\sin\theta\ket{\mbf{e}_2}$ and encoding strategies consisting of only extremal NPE operations (Eq.~\eqref{eq:NPE-Kraus-operators}). With this simplification, we only need to optimize the Holevo information over cq states $\sum_x p_x\state{x}\otimes\rho_x$ where $p_x$ is the prior probability and
\begin{align}
    \rho_x =
    \begin{pmatrix}
        \gamma_x\cos^2\theta & & \\
         & (1-\gamma_x)\cos^2\theta & \sqrt{1-\gamma_x}\cos\theta\sin\theta e^{i\phi_x} \\
         & \sqrt{1-\gamma_x}\cos\theta\sin\theta e^{-i\phi_x} & \sin^2\theta
    \end{pmatrix}.
\end{align}
This means that
\begin{align}
    \sum_x p_x\rho_x = 
    \begin{pmatrix}
        \sum_x p_x\gamma_x\cos^2\theta & & \\
         & \sum_x p_x(1-\gamma_x)\cos^2\theta & \sum_x p_x\sqrt{1-\gamma_x}\cos\theta\sin\theta e^{i\phi_x} \\
         & \sum_x p_x\sqrt{1-\gamma_x}\cos\theta\sin\theta e^{-i\phi_x} & \sin^2\theta
    \end{pmatrix},
\end{align}
and
\begin{align}
    \sum_x p_x S(\rho_x) = \sum_x p_x h_2(\gamma_x\cos^2\theta).
\end{align}
Therefore,
\begin{align}
    \chi\left(\sum_x p_x\state{x}\otimes\rho_x\right) &= S\left(\sum_x p_x\rho_x\right) - \sum_x p_xS(\rho_x) \leq \tilde{\chi}(\theta;p_x,\gamma_x)
\end{align}
where
\begin{align}
    \tilde{\chi}(\theta;p_x,\gamma_x) \coloneqq H\left(\left\{\sum_x p_x\gamma_x\cos^2\theta,\;\sum_x p_x(1-\gamma_x)\cos^2\theta,\;\sin^2\theta\right\}\right) - \sum_x p_x h_2(\gamma_x\cos^2\theta).
\end{align}
Here $H(\cdot)$ is the Shannon entropy, and the inequality becomes equality when $\sum_x p_x\sqrt{1-\gamma_x}e^{i\phi_x}=0$. Taking derivative of $\tilde{\chi}$ with respect to $\gamma_x$, and after some algebra, we find 
\begin{align}
    \frac{d\tilde{\chi}}{d\gamma_x} = p_x\cos^2\theta\left(\log\frac{\sum_{x'} p_{x'}(1-\gamma_{x'})}{\sum_{x'} p_{x'}\gamma_{x'}}-\log\frac{1-\gamma_x\cos^2\theta}{\gamma_x\cos^2\theta}\right)
\end{align}
If there is a local max, then $\displaystyle\frac{d\tilde{\chi}}{d\gamma_x}=0$ for all $a$, which means $\gamma_x = \displaystyle\frac{\sum_{x'} p_{x'}\gamma_{x'}}{\cos^2\theta}\;\forall x$, i.e., they are all equal. However, this then means $\gamma_x=\displaystyle\frac{\gamma_x}{\cos^2\theta}$, which cannot be true unless $\cos^2\theta=1$. Therefore, if $\cos^2\theta\neq 1$, then $\tilde{\chi}$ has no local extrema, and the maximum has to occur at the boundaries $\gamma_x=0$ or $\gamma_x=1$, corresponding to phase shift or complete damping encoding operations.

Thus to maximize $\tilde{\chi}(\theta;p_x,\gamma_x)$ with respect to $\gamma_x$, $\gamma_x$ must be 0 or 1. In this case, let us define $p\coloneqq\sum_{x:\gamma_x=1}p_x$. Then we have
\begin{align}
    \tilde{\chi}(\theta;p_x,\gamma_x) &\leq H\left(p\cos^2\theta,(1-p)\cos^2\theta,\sin^2\theta\right)-ph_2(\cos^2\theta) \notag\\
    &=\cos^2\theta h_2(1-p)+(1-p)h_2(\cos^2\theta)
\end{align}
which means that
\begin{align}
    \chi(\mc{Q}_1^\text{ass}) &\coloneqq \max\chi\left(\sum_x p_x\state{x}\otimes\rho_x\right) \notag\\
    &\leq \max_{p,\theta} \cos^2\theta h_2(p)+ph_2(\cos^2\theta) \notag\\
    &= \max_{x,y\in[0,1]} xh_2(y)+yh_2(x)
\end{align}
Note that this upper bound can be achieved by precisely the encoding scheme given in Eq.~\eqref{Eq:one-sender-encoding}. 

It remains to show that 
\begin{align}
    \max_{x,y\in[0,1]} xh_2(y)+yh_2(x) = \max_{x\in[0,1]} 2xh_2(x).
\end{align}
First, by the symmetry of the objective function, if $(x^*,y^*)$ maximizes the objective function, then $(y^*,x^*)$ must also maximize it. Moreover, observe that $$(1-x)h_2(y)+yh_2(1-x) > xh_2(y)+yh_2(x)$$ for any $x<1/2$. Therefore, we must have $x^*\geq1/2$, and similarly $y^*\geq1/2$. By Lemma \ref{lemma:f-concave} below, the objective function $xh_2(y)+yh_2(x)$ is concave in $[1/2,1]\times[1/2,1]$. Therefore, $\left(\frac{x^*+y^*}{2},\frac{x^*+y^*}{2}\right)$ must be a maximum, too.

To solve $\max_{x\in[0,1]} 2xh_2(x)$, we take the derivative of the objective function and set it to 0, giving us a transcendental equation $h_2(x^*)+x^*\log\left(\frac{1-x^*}{x^*}\right)=0$. Solving this equation numerically yields $x^*\approx0.7035$, at which point the objective function takes the maximal value $1.2339$. In other words, the optimal initial state is $\ket{\psi_{init}}\approx\sqrt{0.7035}\ket{\mbf{e}_1}+\sqrt{0.2965}\ket{\mbf{e}_2}$, and the optimal encoding is given by Eq.~\eqref{Eq:one-sender-encoding} with optimal prior probabilities $p(x)=(1-x^*,x^*/2,x^*/2)\approx(0.2965,0.3518,0.3518)$. 
\end{proof}

\begin{lemma}
\label{lemma:f-concave}
    The function $f(x,y)\coloneqq xh_2(y)+yh_2(x)$ is concave in the region $(x,y)\in[1/2,1]\times[1/2,1]$.
\end{lemma}
\begin{proof}
    The Hessian of $f(x,y)$ is
\begin{align}
    H = 
    \begin{pmatrix}
        \partial^2 f/\partial x^2 & \partial^2 f/\partial x\partial y \\
        \partial^2 f/\partial y\partial x & \partial^2 f/\partial y^2
    \end{pmatrix}
    = \frac{1}{\ln2}
    \begin{pmatrix}
        -\frac{y}{x(1-x)} & \ln\left(\frac{(1-x)(1-y)}{xy}\right) \\
        \ln\left(\frac{(1-x)(1-y)}{xy}\right) & -\frac{x}{y(1-y)}
    \end{pmatrix}.
\end{align}
Calculating the eigenvalues of $H$ reveals that $H\leq0$ if and only if
\begin{align}
    \frac{1}{(1-x)(1-y)}-\left[\ln\left(\frac{(1-x)(1-y)}{xy}\right)\right]^2 \geq 0.
\end{align}
This is true for all $(x,y)\in[1/2,1]\times[1/2,1]$ since
\begin{align}
    \sqrt{\frac{1}{(1-x)(1-y)}} &\geq \ln\left(1+\frac{1}{(1-x)(1-y)}\right) \notag\\
    &\geq \ln\left(1+\frac{x+y-1}{(1-x)(1-y)}\right) \notag\\
    &= \ln\left(\frac{xy}{(1-x)(1-y)}\right)
\end{align}
for any $x,y\in[1/2,1)$. Since the Hessian $H\leq0$ for all $(x,y)\in[1/2,1]\times[1/2,1]$, the function is concave in this region.
\end{proof}
\begin{comment}
    
\textcolor{red}{[AW] Slightly different argument as to why $\det H \geq 0$: If $xy =: u \geq a^2$, then $(1-x)(1-y)=1-x-y+xy =: t \geq (1-a)^2$. Then we argue that for $a=1/2$,
\[
  \frac{1}{t}-\left(\ln\frac{t}{u}\right)^2 \geq \frac{1}{t}-\left(\ln t\right)^2 \geq 0,
\]
because $t\leq \frac{1}{4} \leq u$.
}

\end{comment}

\section{Details of the experiment}

\textbf{Source preparation}:  The source of photon pairs is based on type II spontaneous parametric down-conversion in a 2 mm periodically polled Potassium titanyl phosphate (PPKTP) crystal (with temperature stabilizing oven). The crystal is pumped with frequency-doubled light pulses originating from a Tsunami modelocked laser (a train of $\sim$100-fs pulses with center wavelength 810 nm and repetition rate 80 MHz), doubled using a 0.5 mm Bismuth Borate (BiBO) crystal. To prepare the photons in a single spectral, polarization, and spatial mode, the heralding photons from the pair are filtered to $\sim$2~nm bandwidth at full-width-at-half-maximum by a pair of tilted spectral filters, set to linear polarization by a polarizer, and coupled into single-mode fiber. The existence of this idler photon is detected via a single-photon detector (avalanche photodiode, Excelitas SPCM-AQ4C), while the other, heralded single photon is sent to a three-port interferometer to prepare the desired state $ \ket{\psi}=\frac{1}{\sqrt{2}}\ket{\mbf{e}_1}+\frac{1}{2}\ket{\mbf{e}_2} +\frac{1}{2}\ket{\mbf{e}_3}$, where we have ignored all the internal degrees of freedom of the single particle and only represented it in a superposition of different path basis states $\ket{\mbf{e}_i} = \ket{0}^\msf{A_1}\cdots\ket{1}^{\msf{A}_i}\cdots\ket{0}^{\msf{A}_N}$.
\par
\begin{figure}
    \centering
    \includegraphics[width=0.99\textwidth]{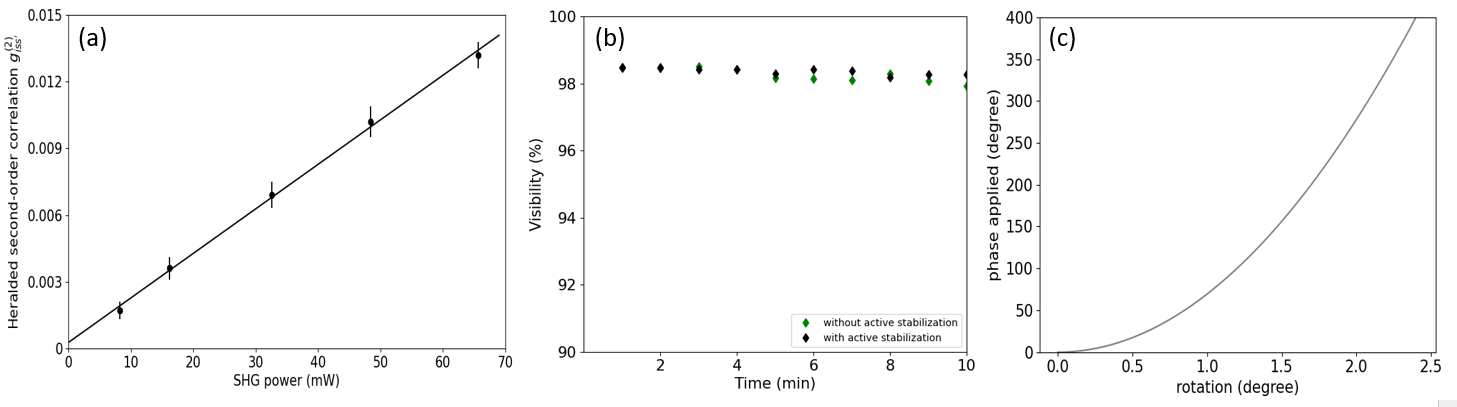}
    \caption{(a) Experimental result for heralded second-order correlation $g_{iss'}^{(2)}$ under different SHG pump powers; (b) Interference visibility of the Mach–Zehnder interferometer measured in 10 minutes. (c) Theoretical characterization of our phase plate.}
    \label{fig:experimentdetail}
\end{figure}

 
To ensure the signal photon is close to a single photon source, we looked at the heralded signal photon within a $2$ ns coincident window after heralding the idler photon. We characterize the source by measuring its second-order correlation $g_{iss'}^{(2)}$, which can be calculated as:
\begin{equation}
g_{iss'}^{(2)}=\frac{C_{iss'}C_i}{C_{is}C_{is'}}
\end{equation}
where $C_{iss'}$ are three-fold coincident counts between one idler photon and two single photons after splitting, $C_{is(s')}$ represents two-fold coincident counts between idler photon and one signal photon, and $C_i$ denotes single counts for the idler. The power-dependence of the second-order correlation values is shown in Fig.~\ref{fig:experimentdetail} (a), which indicates good agreement with the linear curve fitting and  allows to set the pump power to suppress the two-photon contribution from the source. \par
In our experiment, the heralded second-order correlation $g_{iss'}^{(2)}$ has to be set extremely small, due to the fact that large higher-order terms could in principle enable a higher capacity rate even in the classical case. Taking the small violation we have estimated ($1.02$), we set $g_{iss'}^{(2)}(0)$ = $0.0017 \pm 0.001$ to be one order of magnitude smaller than the violation to make sure the contribution from multiple-photon events can be neglected. As a consequence, we have relatively low coincidence count rates around 600Hz
\\
\par 
\noindent \textbf{Interferometer design}: The interference visibility of our three-port interferometer limits the performance of our quantum-enhanced communication. To achieve a high enough visibility with free-space optics, we design a three-port interferometer consisting of (1) an inner offset Sagnac interferometer, which is extremely stable over a few hours with above $99.5\%$ interference visibility; (2) an outer Mach–Zehnder interferometer, which is passively stabilized thermally and vibrationally inside a small box and gives around $98.2\%$ interference visibility over 10 minutes. It was further actively adjusted by a piezo actuator implemented on the translation stage in the delay line between different runs of measurement; (3) three $3$-mm-thin glasses windows for controlling the phase independently; windows were chosen instead of other bulky electro-optical devices, which could potentially degrade the interference visibility.
\par
The whole setup can be maintained stable over $\sim$10 minutes with average interference visibilities around $99.5\%$ and $98.2\%$ for the inner and outer loops, respectively, while slight adjustment with the piezo actuator helps to retrieve good interference visibility for the next round of the experiment. During the runs of our experiment, we do not turn the active stabilization on so that the average stability remains the same over 10 minutes. \\
\par
\noindent \textbf{Encoding operation}. As has been mentioned before, with the current type of single-photon detectors used and the loss in our free-optics setup, performing general amplitude damping operations on the photons is nontrivial. Instead, we devised our setup based on the coherent-assisted protocol where only phase encoding is required. \par
One of the most commonly used phase shifters is electrically controlled liquid crystal, where the refractive index along some axes of the crystal depends on the voltage applied to it and thus can be used to add phase on single photons. However, the resolution of the applied phase (around $3^{\circ}$) and the size and the parallelism of most commercial liquid crystals prevent us from using them in our small-size, high-visibility interferometer.  Therefore, as a replacement, we create a phase shifter based on a $d=3$ mm glass window (with a reflective index around $n_g=1.51$) mounted on a rotation stage (with a resolution around $25$ second-arc). Starting from placing the glass plate perpendicular to the incoming beam, the phase added to the photon after slightly tilting it with angle $\alpha$ can be computed as:
\begin{equation}
\Delta\phi = \frac{2\pi d}{\lambda}[(\sqrt{n_g^2-\sin(\alpha)^2}-\cos(\alpha))-(n_g-1)],
\end{equation}
which is plotted in Fig.~\ref{fig:experimentdetail}. The average resolution over $2\pi$ phase shift is around $2^{\circ}$; however, due to its nonlinear behavior, by carefully choosing the starting point, we can obtain a much finer resolution. 
\\
\par
\noindent \textbf{Error analysis}: To estimate the experimental error, we note at first that we are limited mostly by the interferometer stability. To ensure high interference visibility, we perform each run of our measurement for $\sim$10 minutes and re-optimize the setup between different runs.
\par
In each run of the experiment, the statistical error can be calculated from standard error propagation. For the case of characterizing channel transition probability $p(y|x)$:
\begin{align}
V[R_1]&=\frac{1}{N^2}\sum_{xy}\left[\left(p(x)\log_2 q(y)+p(x)H(q(y))\right)^2+\left(p(x)\log_2  p(y|x)+p(x)H(p(y|x))\right)^2)\right] V[n_{y|x}]
\end{align}
where $q(y)=\sum_x p(y|x)p(x)$ with fixed optimal prior $p(x_1=0)=1/2$ and $p(x_2=0)=15/17$. $n_{y|x}$ is the total number of photons collected at port $y$ conditional on input $x$ and $N$ is a total number of counts using in characterizing the channel for every input $x$. The statistical error assuming the Poisson distribution is given as $V(n_{y|x})=Np(y|x)(1-p(y|x))$. With $N\approx 10^5$, the statistical error is around $\sqrt{V(R_1)}\approx 0.002$.\par 
Similarly, for the case of measuring the joint distribution $p(a,b)$, the error can be computed as:
\begin{equation}
V[R_2]=\frac{1}{N^2}\sum_{x,y}\left(\log_2q(y)+\log_2 p(y))-\log_2 p(x,y)+I(y:x)\right)^2 V[n_{xy}],
\end{equation}
With extra uncertainty in $p(x)$ from the generation of encoding random bits and $V[n_{xy}]=np(x)(1-p(x))\times m+n\times mp(y|x)(1-p(y|x))$ where $n=680$ (the number of random input bits) and $m\approx 600$ (the number of counts per second), we get a statistical error of estimating $V[R_2]$ to be around $\sqrt{V(R_2)}\approx 0.011$. \par 
However, experimentally, besides the statistical error, the channels built from run to run are actually slightly different since they are extremely sensitive to the overall interference visibility. To take those systematic errors into consideration and to show that our experimental result is repeatable, we calculate the experimental result $I(X: Y)$ in 10 runs of experiments and average the capacity rate sum, which is what we present in the main text.   

\end{document}